# Report

# on

*Development of strategies to improve*
*the students' learning*
*by*
*participation in class*
*and*
*out-of-classroom.*

January 2011

Rob J.W.E. Lahaye & Natanael Karjanto

# Contents

# Introduction

Education is the part of one's life in which knowledge, skills, understanding, and moral values are acquired. According to Wikipedia, an online encyclopedia, education is the process by which people learn instruction, teaching and learning.

Many governments throughout the world are concerned with the quality of elementary, secondary, and higher education of their citizenry. International standards are developed in order to verify and compare the effects of education policies. Examples of assessments in reading, mathematics, and science literacy are the Progress in International Reading Literacy Study (PIRLS), the Trends in International Mathematics and Science Study (TIMSS), and Programme for International Student Assessment (PISA). The first two studies were conducted by the International Association for the Evaluation of Educational Achievement (IEA) in order to compare students' educational achievements across borders among the participating nations; the latter one is coordinated by the Organization for Economic Cooperation and Development (OECD) aiming at improving the educational policies and outcomes. The results of the assessments can help educational policy makers improve education strategies. Although the assessments are predominantly for the elementary and secondary school levels, by no doubt this also reflects on the quality of the higher education. In particular for higher and college education, many institutions worldwide develop programs (or "visions") to improve educational quality through teaching and learning.

A lot of resources have been invested into programs to improve the quality of education by training instructors, changing curricula, and introducing new and innovative styles in teaching. Although the conventional-traditional teaching method remains popular around the globe, pioneer-minded institutions challenge these teaching methods by successfully implementing up-to-date approaches in teaching and instruction. Examples are problem-based learning and guided-discovery, which are both based on constructivist teaching strategies. Active learning is a method which changes the balance between the instructor's role and that of the students.

There is an abundance of literature on active learning, stimulating active participation, and enhancing critical thinking. In this context, active learning refers to a collection of techniques where students do more than simply listening to an instructor, but also discover, process, and apply information themselves (McKinney, 2010). According to Meyers and Jones (Meyers&Jones, 1993), active learning derives from two basic assumptions: learning by nature is an active endeavor, and different people learn in different ways. The authors also show that the quantity of learning is enhanced when the students are engaged in active learning. Bonwell and Eison popularized the approach of active learning for class instruction (Bonwell&Eison, 1991). They suggested that pupils can work in pairs, discuss the materials while role-playing, debate, engage in case study, take part in cooperative learning, produce short written exercises, and many other activities. Lazarus mentions a number of strategies to stimulate active participation, for instance by questioning techniques and stimulating group discussions (Lazarus, 1999). Kim and Kim explore students' participation in online discussions (Kim&Kim, 2006). The authors discover that the participation patterns are mainly influenced by the students' voluntary participation. Recently Salter emphasized that the implementation of problem-based learning techniques can change surface learning into deep learning, memorized knowledge into understanding, etc. (Salter, 2010). Tang and Titus showed the positive effects of incorporating interactive and media-enhanced lectures to promote active learning in Calculus and Physics courses (Tang&Titus, 2002). Also Orhun and Orhun studied the relation between learning styles and achievements in Physics and Calculus courses (Orhun&Orhun, 2006). Marongelle presented active learning in using a ‘studio’ approach by combining the Calculus and Physics course (Marongelle, 2003).

The focus of our study and this report is the results of a search for improving our own teaching skills and analyze the effects on and responses of the students. This has led us to interesting conclusions, though this is not the end of our search for teaching improvements. This report is about the intermediate status of what probably is a life-long study into becoming a better instructor. We endeavour to improve the students' learning skills both inside and out-of-classroom. We hope that students not only improve their learning skills in the subjects we teach, but also elevate their learning achievements in other subjects later. With this in mind, it is expected that what we strive to implement would give a beneficial effect to the students, both in the short term and in the long term period.

Within the limitation of the subject we teach, our main interest is to enlarge and stimulate the quantity of the students' learning for the Calculus and General Physics courses offered as the Basic Sciences and Mathematics (BSM) modules in the Natural Sciences Campus of Sungkyunkwan University (SKKU). Increasing the quantity of learning requires new teaching strategies and different approaches as to how students acquire knowledge. We therefore investigate and implement some of the aforementioned approaches including group assignments, guided-discovery and English delivery lectures for an audience, whose first language is not English.

In this report we discuss the findings we have observed in the Spring and Fall semesters of 2010 for the General Physics 1 and 2 courses, and in the Fall and Winter semesters of 2010 for the Calculus 2 course. The General Physics and Calculus courses are different in nature and are offered independently by two instructors from different departments with different specialties in educational backgrounds. Nevertheless, these two courses are closely related in terms of logical development and critical understanding. In both courses we aim at stimulating students' active participation and improving the teaching method that fits best to the group of students.

Finally, a philosophical note on the importance for instructors to be continuously alert for teaching improvements. Generally, instructors tend to teach using their own educational, cultural, economic, and moral baggage. As a result, instructors more often than not get trapped in a teaching style similar to the one in which they were taught, giving rise to a generation gap in teaching development. A significant change in education is only possible when instructors are actively involved with up-to-date teaching methods and are given the opportunities to learn and develop new teaching skills. Moreover many effectual contemporary teaching techniques have been proven to be best suited to the present-day students. A few students in our classroom will pursue an academic carrier later in their professional lives. We should therefore also see the classroom as the breeding ground for the next generation academic instructors. Setting good examples by current instructors helps them to become better instructors in the future.

This report is organized as follows. At first, we discuss the findings for the General Physics course, followed by the findings for the Calculus course. Finally, in the conclusion we summarize the approaches in the two different courses and also briefly look at possible future improvements.

# General Physics Course

## In the classroom

### Course management

General Physics is divided into two separate courses, one in each regular semester. General Physics 1 in the Spring semester (code GEDB008) focuses on classical mechanics and thermodynamics, whereas General Physics 2 in the Fall semester (code GEDB010) deals with electro-magnetism and modern physics. The courses are scheduled every year. Taking the two courses in chronological order is recommended, but since the university administration manages the two courses independently, students can as well sign up for the two courses in reversed order.

Since both General Physics courses are mandatory for science liberal arts students and for most science major undergraduates in the Suwon campus of SKKU, every semester approximately 1000 students register for the course and 10 to 12 instructors teach some 15 to 20 different General Physics classes. Students can choose the class according to their preferences for class hours and instructor on a first come-first-served basis, provided the class of choice has not yet reached the maximum of 70 registered students. Due to the large number of students, classes are usually filled to their maximum capacity. Only during the very first week of the course students are allowed to withdraw from the course or switch to another class.

The students in all General Physics classes take the same standard midterm and final exams. Therefore the textbook, study topics, and objectives are standardized by the physics department. The instructors are given a detailed course description, which contains per week the chapters to teach and the chapter-parts to be skipped. Instructors have a measure of freedom to choose their own teaching style and assignment policy, as long as the students are adequately prepared for the standard midterm and final exams.

Per week the students spend two hours and 30 minutes in the classroom. Depending on the class, this is split into two parts of one hour and 15 minutes each on different days, or as one block of two and half hours.

## Textbook and syllabus

There is an abundance of textbooks suitable for an American college style General Physics course. All textbooks cover more or less the same topics and differ only in details. The physics department has picked an English language textbook (Jewett&Serway, 2008) on which the syllabus is based. All instructors get a free copy of the textbook and in the campus bookstore the students can purchase a fairly cheap student edition. Although the students are advised to use this particular textbook, they can also purchase an alternative textbook or a Korean translation of the corresponding textbook; students then need to carefully keep track of the chapter correspondences between their own book and the official textbook. It is important to emphasize to students with a Korean textbook that they also need to put extra efforts in making themselves familiar with the English terminologies of the physics concepts studied in the course.

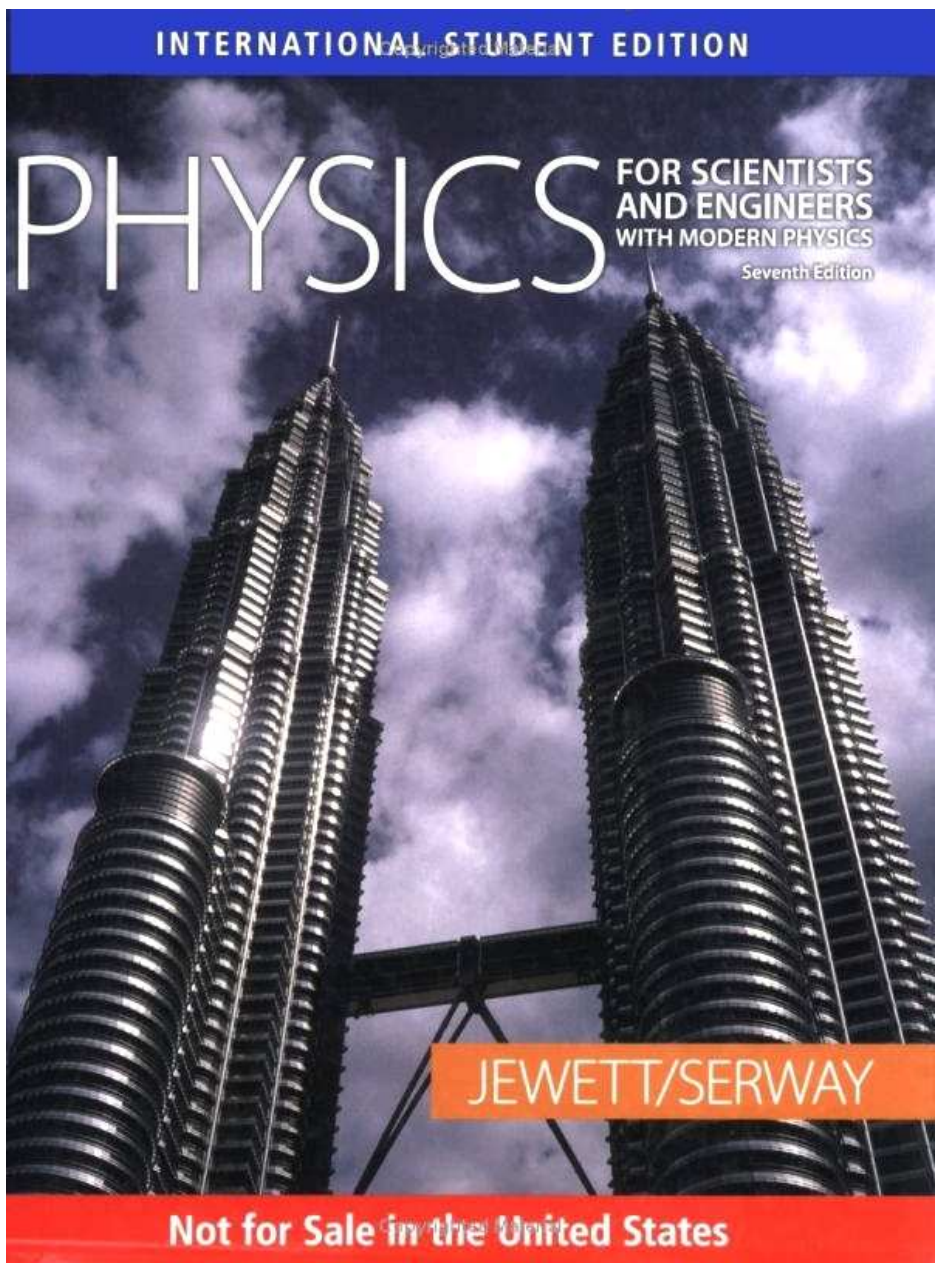

*The cover of the textbook for General Physics 1 and 2.*

When the students register for a particular class before the semester starts, they already have access to the syllabus and course objectives through the ASIS-GLS information system, which also provides information on the grade and evaluation policy.

## Implemented teaching method

The authors of the physics textbook have prepared a CD-ROM exclusively for the instructors of the course and its content is supposed to ease the teaching load of the instructor. It contains power-point presentations for each chapter of the book and a solution manual to all problems in the textbook. The suggested teaching method implies the guidance of the students through the power-point presentations during the lectures and give assignments to the students from the textbook problems.

The advantage of power-point oriented lectures is that they are time efficient. Instructors often rely on this teaching style when there seems to be not enough time for dealing with all topics of the syllabus, or when lecturing time needs to be reduced in favor of quiz or problem-solving sessions. Most students, on the other hand, perceive power-point-style lectures as boring and their minds very quickly drift away from the presentation and the instructor. This leads to the paradoxical conclusion that a power-point based teaching style is doomed to be inefficient, as students 'waste' a lot of quality learning time in class.

Instructors are not allowed to share the solution manual on the CD-ROM with the students, because this would infringe the agreement with the printing company of the textbook. This kind of practice can be justified within a high school environment, but it is not a good example of academic learning. Students should not be trained to find the one and only correct solution, which is hidden from them by the instructor. Moreover, some students are smart enough to find ways to get hold of 'illegal' copies of the solution manuals, which gives them an advantage to others who do not have them. In a competitive environment this can be harmful and trigger wrong attitudes among the students.

In order to avoid the risks and traps with the power-point presentations and the solution manuals on the CD-ROM, I decide not to use this as my teaching materials, but instead create my own teaching materials derived from the pictures in the textbook and tailored to my own story of the lecture. This is further supported by animations and movies I found elsewhere to help students better understand new concepts. All my teaching materials are available online as a website from a web-server, which runs on my office computer. As the website is always online, students have access to the materials before, after, or even during class (for those using a wired laptop in class), while I myself download the same material for my lectures to the classroom computer. In the lecture I project everything on the whiteboard, so that remarks or additional explanations can be added into the projection with the whiteboard markers. This is convenient when for example I want to manually add more complexity into an initially simplified picture.

Although using an interesting lecturing style helps keep the students' concentrated, most students would gradually lose focus anyway after 15 to 20 minutes into the lecture and students' learning efficiency decreases rapidly after this time span. A variety of techniques can reset or revive the students' concentration. For example an entertainment intermezzo is often effective, such as showing a short movie or telling a joke. Alternatively, one can introduce classroom activities, in which the students must change from a passive audience into active participants. Alternating this with traditional lecture style prevents students from dozing away during class time.

In order to find an appropriate balance between lecturing and classroom activities, I prepare a lecture plan. For one hour teaching I write out the list of topics to be lectured on an A4 paper and in between the topics I add questions for the classroom activities. If it looks well balanced on paper, it most often also works fine in the classroom.

When I did class sessions without problem-solving activities, I then used movies or animations in between the lecturing times to keep students alert until the end of the class. I noticed that students find this kind of physics entertainment less fun than the classroom activities. The students seem to appreciate it more if they can switch from passive listeners or viewers to group work which allows them to talk to their fellow students while I am walking around to answer their questions. The social fun-factor and the personal contact with the instructor is apparently more important than the entertainment fun-factor.

Occasionally one student or a few disturb the class by being engaged with noisy activities which have nothing to do with the ongoing lecture. Most often it is sufficient to politely ask the students to be quiet. However, every now and then, the same students continue to disturb the class again after a little while and it becomes a continuous distraction for both instructor and fellow students. Appropriate measures are then needed to stop the students, without interrupting too much the lecture and the class atmosphere. Once I stopped lecturing for a moment, scolded the noisy students, and asked them to leave the classroom. Although this solved the problem, it created an awkward scene in class. Afterwards I felt very uncomfortable and the class atmosphere remained icy after the incident. I concluded that this is not a good approach to deal with the problem. It is preferable to address such students privately after class or during break-time and then explain to them that irresponsible behavior in class is not accepted. This also gives the opportunity to the students to explain why they were noisy and sometimes a mutual understanding of the situation can settle the issue more peacefully.

## Attendance record and students' participation

There is an apparent rule in SKKU that instructors must take attendance records of every lecture. For the large number of students in the General Physics classes it is not practical to follow this rule strictly and some instructors compromise on fairness by calling the names of a few randomly picked students at the beginning of the class. In my opinion either all students should be checked, or no check at all. Calling all names and verify the person with the photograph in the list is too time consuming. Therefore I have chosen to integrate the attendance check with the classroom activities, which guarantees a complete and fair attendance record.

The General Physics courses during the Spring and Fall semesters are in room 62354, which has separate desks and chairs for each individual student. The common arrangement for the desks is pairwise in rows with all students facing the instructor and the whiteboard. For the classroom problem-solving activities, the students must rearrange the desks into groups of four, in such a way that it is still possible for all students to comfortably watch the whiteboard. To each group I hand out one A5 sized paper. The paper has four fields for the names and student numbers and the rest is blank, where the students should write the group's answers. During the group activities students have to discuss and find answers to questions and I walk around the classroom. I then verify that the names on the papers match with the size of the group, to prevent students squeezing in the names of their absent friends. At the end of the class time I collect the papers again and use them for the attendance record. The advantages of this type of attendance check is that it is fair and does not waste time. After class it takes about 15 minutes to enter the attendance data of 60 to 70 students into the computer.

| Week | General Physics 1<br>Class 54<br>72 students | General Physics 1<br>Class 59<br>72 students | General Physics 2<br>Class 41<br>70 students | General Physics 2<br>Class 52<br>70 students |
|---|---|---|---|---|
| 1 | 69 | 65 | X | X |
| 2 | 71 | 71 | 53 | 53 |
| 3 | 72 | 69 | X | 57 |
| 4 | 69 | 69 | Chuseok vacation | Chuseok vacation |
| 5 | 65 | 67 | X | X |
| 6 | 63 | 63 | 38 | 53 |
| 7 | 65 | 59 | 051 | 46 |
| 8 | Midterm exam | Midterm exam | Midterm exam | Midterm exam |
| 9 | X | X | 24 | 46 |
| 10 | 63 | 53 | 18 | 37 |
| 11 | X | X | 31 | 36 |
| 12 | X | X | 24 | 41 |
| 13 | X | X | 31 | 32 |
| 14 | X | X | 33 | 30 |
| 15 | 43 | 40 | X | 32 |
| 16 | Final exam | Final exam | Final exam | Final exam |

*Attendance counts per week for the General Physics 1 courses in the 2010 Spring semester and for the General Physics 2 courses in the 2010 Fall semester. X means attendance was not checked.*

The attendance check from the classroom activities answer papers also shows information on the attendance fluctuations from week to week during the whole semester, that is for the weeks that I checked attendance. It is difficult to fully understand these fluctuations, because students may have their private motivations when to attend or skip a class. The trend seems to indicate that students find it easier to skip classes after the midterm exam when the semester end is approaching. For some students the midterm exam is the litmus test of whether it is worth to spend more time on General Physics and get a good grade, or give up on it altogether and focus instead on other courses.

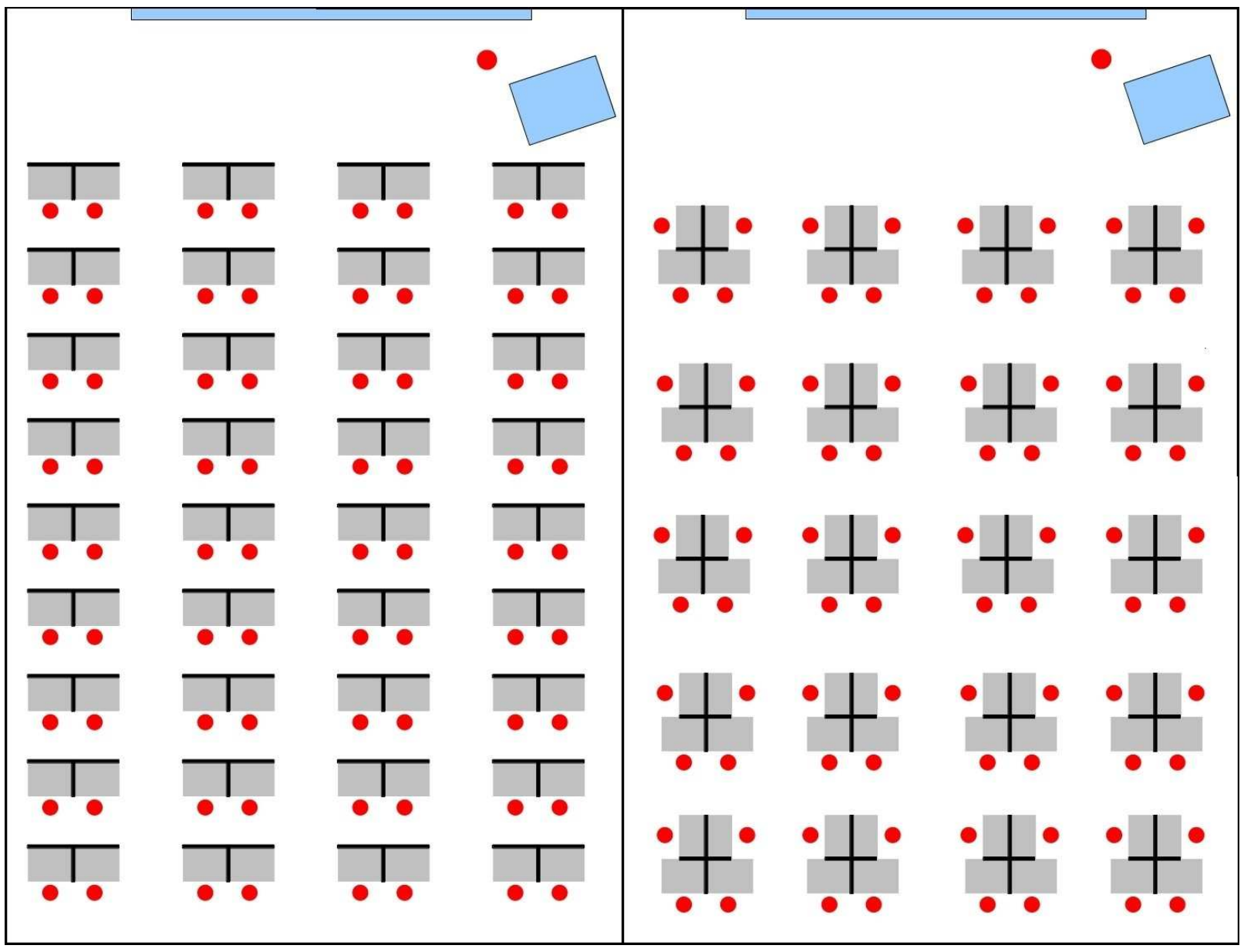

*Left: The standard classroom arrangement with the desks pairwise in rows.*
*Right: The new desk arrangement for group activities in the classroom.*

The new classroom arrangement provides the basis and framework for the group activities. The group formations in class are allowed to vary from class to class and thus the students can form groups ad hoc every time they come to class. However, as the semester evolves, about half of the students spontaneously team up with the same group members each class. These are the students who also have a constant attendance record. The group work allows and stimulates students to talk to each other and hence makes the time spent in class a more social event. The students appreciate to not only attend class in order to quietly listen to the instructor's teaching, but also participate in problem solving sessions together with the students in their own group. The students are asked one or more questions related to the topics discussed in the lecture session prior to the group work. In order to stimulate the group activity, I walk around the classroom to help students and answer additional questions in private discussions with the groups. This often leads to intensive team work, because each group needs to reach a consensus on their final answers.

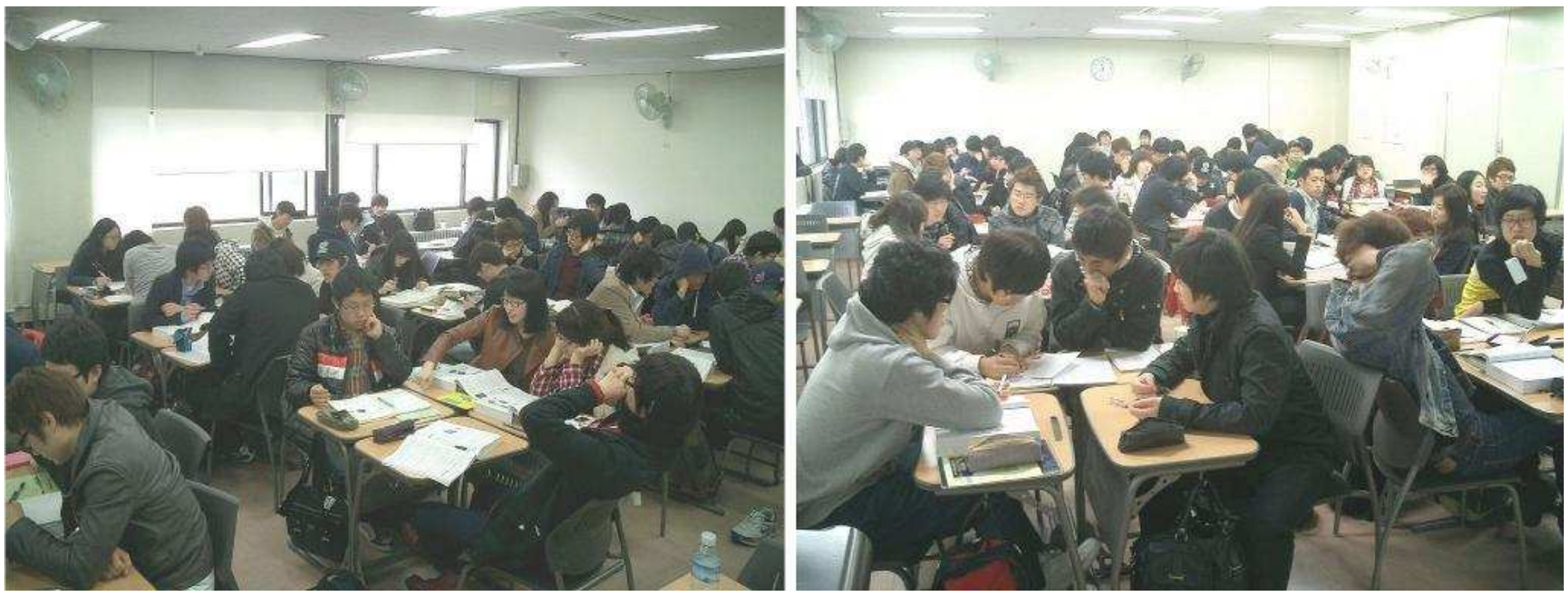

*In the classroom activities students discuss their answers to physics questions.*

During the group activity sessions students themselves become learners and teachers at the same time, because they have to share and explain their understanding of a topic to the other group members, convince them of their points of view, and listen to what others have to say. These group discussions help students verify whether their understanding is sufficient to solve real problems and compare their own understanding with that of others in the group. When this type of student interaction works well, it can be a far more effective learning time for the students than being lectured by the instructor.

A group activity in the classroom typically lasts for about 15 minutes. At the end of the group discussion the group has to reach consensus on the answers and write them on an A5 paper given to them at the beginning of the class. During a one hour lecture there can be one or two such sessions. I will collect the answer papers of all groups at the end of the class. These papers are then used for attendance records and occasionally students get extra points for their answers.

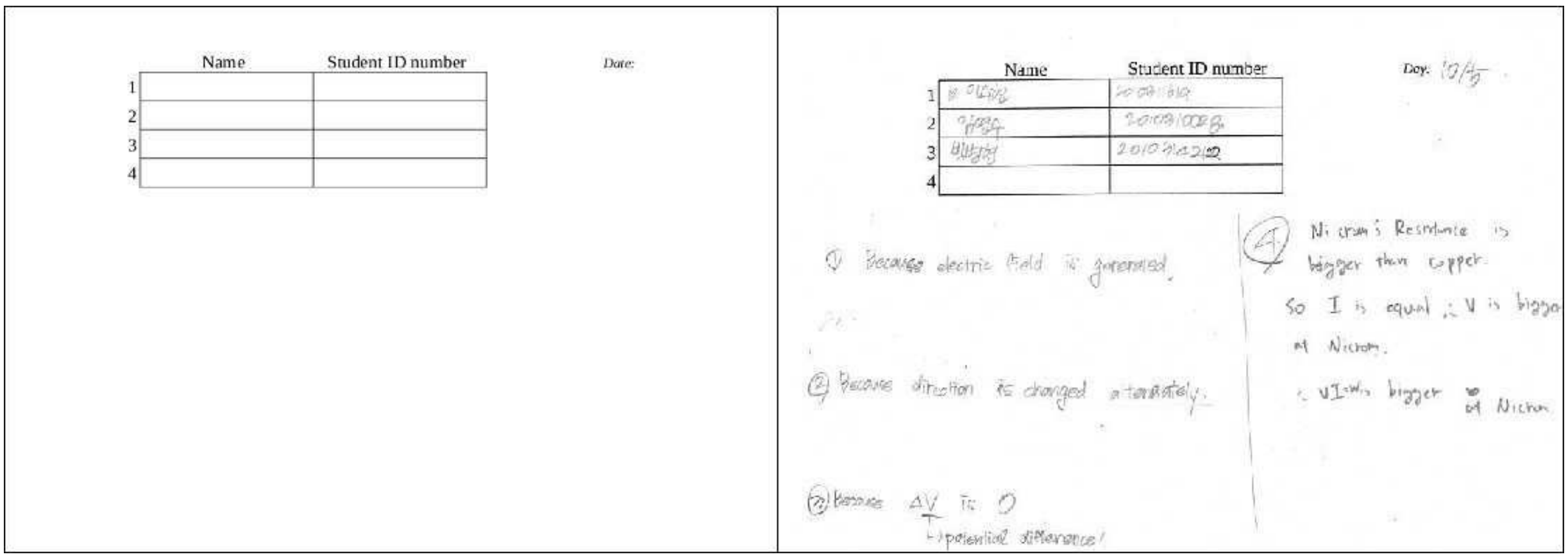

*A5 size answer paper for quizzes as group activities in class, handed out one per group at the beginning of the class (left) and collected at the end of the class (right).*

There are occasions when students resist the group work and indicate preference for solving the problems by themselves. It is a dilemma whether or not to enforce group work onto these students. Should it be mandatory to work in groups in the classroom? I have decided to be lenient towards students and allow them to work alone if their arguments for working alone are convincing. My experience is that these students then are highly motivated to work on and solve problems by themselves.

In a questionnaire on all students of the two General Physics 2 courses in Fall 2010, the response shows that many students enjoy the group activities in class: 43.2% said they liked it, whereas 16.8% disliked it, and the remaining 40% were indifferent. The large number of indifferent students implies that they have doubts about the group work. For me this means that the organization of the group activities still needs improvement. For example, I need to be more careful in how to formulate the questions, due to the use of English language, otherwise some students may not get the essence of the question and have no clue what to do; occasionally I found a whole team flabbergasted and completely lost. Another issue may relate to too obscure evidence of how the classroom participation contributes to a better grade. Simply collecting the answer papers at the end of the class does not stimulate students enough to take the classroom activities seriously. If no explicit rules are set for classroom participation, students' motivation to put effort in the group work deteriorates as weeks pass by.

Later in the semester I made an attempt to fight the 'group work deterioration' and tried to tune up the energy of the group work by asking each group to let a representative explain their answers in front of the class. This turned out to be too embarrassing for the students and most presentations became unclear mumbling stories while the rest of the class got bored. Nobody enjoyed this type of group activity and the effect was contrary to my intentions. I immediately abandoned this strategy.

A better solution to keep the spirit high of the classroom group work is a thorough set of assessment rules for the classroom participation, so that the students immediately understand the reward for their efforts. Unfortunately I have not yet figured out how to implement a set of rules to accomplish this. One strategy could be to revamp the teaching method into one that is more centered around the classroom participation, such as "problem-based learning".

A risky side-effect of the group work in class and the furniture rearrangement can be the difficulty to control the unwanted talking and noise at times a quiet classroom is required, for example when it is time for the instructor to lecture at the whiteboard or for watching a movie. Unwanted classroom noise often occurs when students are given insufficient time to finish the group work, or when students have lingered into private talking after given too much discussion time. Hence, as an instructor I should closely monitor the students' progress and keep the students focused on the physics discussion while walking around in the classroom. It is then of vital importance to have one or two supplementary questions ready in case groups are already finished, while others are still busy.

Finally, an anecdote on how group work may need time to become more successful. One of the General Physics 2 lectures dealt with the properties of light. At the end of the lecture I had some spare time left and decided to let students do group work. I had no group activities prepared and thus I had to come up with a problem on the spot. I formulated one simple question: "Design a reliable method to measure the speed of light". This question had no direct links with the lecturing materials of that particular day and many students were at a loss what to do. During the first five minutes or so I myself almost panicked, because very few students seem to know where to start and how to proceed. Fortunately time saved me. While I was walking around the classroom and chatted here and there with some groups, I saw that students were gradually developing ideas and group discussions became more and more active. At the end of some 20 minutes, every group had created their own unique design for measuring the speed of light. The lesson I learned is that effective group work needs time to develop. Even though a question or problem is not well defined initially, if the students are given enough time to brain storm ideas, discuss, and find their own solution strategies, group activities in the classroom can be effective, pleasant, and useful.

# Out-of-classroom

## Icampus

By default, the General Physics courses are plain offline lectures. At the beginning of the semester I usually apply for an upgrade to e+ courses. The classroom lectures can then be recorded, subsequently uploaded to icampus, and students can view the lectures later online. This is helpful to students who have difficulties with the English language in class and want to listen to parts of a lecture again, but also students who accidentally missed a class can catch up online. About one third to half of the students in the course watch the online recordings. Therefore I believe that the e+ recordings are useful to keep students engaged with the lecture materials.

Icampus also offers the tools to communicate online with the students. Besides the one-way announcements by the instructor or the teaching assistants, students can use a Q&A board to ask questions or start a discussion, either with the instructor or with other fellow students in the course. Especially with international faculty, I have noticed that students find it easier to communicate by written messages than visiting the instructor's office. The range of questions in icampus Q&A varies from asking clarifications of a physics topic in class, to students apologizing for having been absent in the class. Students apparently find this a useful means of communication with the instructor.

In the very first class I point out to the students the advantages of using icampus Q&A for discussing non-personal issues related to the General Physics course. If a student asks a physics problem or question in icampus Q&A, then all students can benefit from the discussion and even may join in. If one student has a physics question, then very often this means that many other students have a similar question too. A public Q&A icampus discussion can solve therefore many problems simultaneously, unlike a one-on-one private discussion with the instructor.

Students who are struggling with personal problems should always seek for a private conversation with the instructor. The student can ask for a personal meeting during office hours, or use the personal message system of icampus. Both options are used by the students, the more confident ones prefer private meetings with the instructor, whereas others feel more comfortable using the personal messaging system in icampus. If students' personal problems are complicated or beyond my own expertise, I introduce them to the student counselor of the University College.

## Assignments

I usually give weekly assignments related to the topics taught in that week and in the past all students had to hand in their own personal solutions. The grading results of the assignments would then contribute to the final grade of the student. The teaching assistants who graded the assignments discovered frequently students who had copied the solutions from others. To stop this plagiarism attitude both the copiers and the copied ones were punished. Unfortunately it is difficult for the teaching assistants to discover copied assignments when he/she has to grade the solutions of 60 or more students and therefore it is most likely that too many copiers got away with it unnoticed. To warn the students for being punished when found copying the solutions of others, did not really help.

Instead of contending with the students for a fair assignment attitude, I decided for a better approach which would make policing of the copy-cats redundant. Students must team up with 3 or 4 fellow students and hand in one copy of the assignment solutions per team. Students who otherwise might be tempted to copy each other's solutions, now become team members and help each other in the team.

There are no assignments yet during the very first week of the course and thus I can give students two weeks to organize themselves into teams. Icampus Q&A is then the tool for students to find a team or for incomplete teams to 'recruit' more members. This is particularly helpful to those students who have few or none acquaintances in the class. By the end of the team formation period, the teams report their members to me via icampus and I give each team a unique team number.

In a class of 70 students, typically 20 teams are formed with 3, 4, or sometimes 5 members per team. Icampus has an option for managing student teams. This is useful when also the assignments and solutions are handled via icampus, which is not the case in my General Physics courses. The major disadvantage of the icampus team management is that teams cannot be changed once they are formed. Although in principle the assignment teams are fixed for the rest of the semester, occasionally team members contact me during the semester with the request for a team change. Here are a few example requests from my own experience:

1. A team member has given up on the course and does not participate anymore in the team meetings.
2. A team member never shows up in the meetings and the other team members want to replace that person.
3. Two teams lost few of their members (as mentioned in previous two examples) and they want to merge.

Instead of the rigid team manager in icampus, I prefer to help students optimize the team conditions by allowing rearrangements of their teams. Requests for changes in a team do not happen so frequently, but often enough to implement a flexible team management. For that it is much easier if I deal with the team administration myself without using the icampus team facilities.

The web-server, which provides all my teaching materials online, also facilitates the distribution of the assignments. Once a week I upload new assignments and from then onwards students can download it at any time. The assignment teams have to organize their own meetings so that they can hand in the solutions one week later.

I do not interfere with the students team work or how they divide the work among the team members. In my opinion the students have to learn from practice how to function well within a team and cooperate with other team members. Demanding team work from freshmen in their first year may not yet result is a perfect team spirit, but it should contribute to the gradual understanding of the students how to manage and organize the workload in a team and how to behave as a team member. It is impossible to learn team work in one semester. It is a long term process of practice and experiences.

The teaching assistant of the class, who has to grade the assignment solutions, obviously prefers teams to individual assignments because it is significantly less grading work. On the other hand, I find it then reasonable to ask for a fast grading, so that the assignments can be returned within a week. If the time for grading takes too long, students may have forgotten their struggles to find the solutions and lose interest in the grading details. Once a week the students can meet up with the teaching assistant for receiving their assignments back and filing complaints about the grading policy.

## Contact hour with the teaching assistants

The University College organizes a weekly contact hour with teaching assistants for all calculus, physics, chemistry, and biology students. In each discipline the number of students is in the order of 1000. The contact hours are scheduled from 6 to 7pm with each subject on a different weekday. A teaching assistant would be waiting for students in a standard classroom (maximum 80 to 100 seats). The contact hours are not mandatory, neither does attendance give bonus points or is there any effect on the students' grades. The contact hours are organized without consultation of the instructors and there is no active coordination between the ongoing classes and the contact hours. The main purpose of this contact hour is to provide an opportunity for the students to get help in the respective subjects in Korean language by a qualified teaching assistant. However, due to lack of coordination the attendance is extremely low, less than 1 percent. For the physics contact hours, very often none, or only 1 student showed up. Despite the good intentions, this contact hour failed to serve its purpose.

Unlike the previously mentioned contact our for all 1000+ general physics students, I organize a weekly “study hour” for my own students in addition to the regular class times. The teaching assistants of my courses are in charge of this study hour. The teaching assistants return the graded assignments to the students and students can ask questions about previous or new assignments. For practical reasons this study hour is often scheduled in the evenings; for both semesters in 2010 it was on Mondays between 7 and 8pm. The study hour is on a voluntary basis and attendance does not affect the grades of the students. I have two motivations to ask this favor from teaching assistants. For one, in this meeting students can ask questions and discuss physics in Korean language (my teaching assistants are always Korean graduate students with good English skills) and I hope that together with my classes in English this improves the students understanding of physics. Secondly, some students may find it easier to discuss with the teaching assistants than with the instructor because of the lower hierarchical barriers. On average 15 to 20% of the students in my classes attend the weekly study hour. As for the teaching assistant, this study hour can be a great opportunity to develop his/her own education skills.

Twice I have conducted a questionnaire among the students and asked their opinion about the study hour with the teaching assistants. Back in 2009 the results were that 32% of all students attended the study hour regularly, of which 71% found it very useful. In 2010 55% of the students attended the study hour regularly, of which 30% found it useful. From both questionnaires one could conclude that roughly 20% of all students make good use of the study hour. Probably these students have a genuine interest in physics without the need of a grade incentive for coming to the study hour.

## Office hours

In principle students can visit my office at any time, if I am available and I have no other pressing business at that moment. However, in order to prevent students coming to my office in vain, I prefer them to contact me by phone or email and make an appointment first. This is not rigid and appointments can be arranged casually; even calls with the request “May I visit your office now?” are welcome.

Unfortunately the University College has recently decided to enforce office sharing by the college professors. This has negative effects on student visits, because it is not comfortable to have a lengthy discussion with a student while knowing that this disturbs somebody else in the same room. The University College has been persistent in its decision, despite protests from all college professors.

Because of the office sharing I try to limit the visits of students to my office, but without compromising the students’ desire to contact me. I therefore ask students to consider icampus or email as a first means to talk to me. Most often problems can be easily resolved via these written media. If problems persist, students usually do not show any impediments to visit me personally.

The students take a midterm and final exam, which is standardized by the physics department for all 1000+ General Physics students. Immediately after the exam the physics department publishes the exam with solutions on their website. The grading of the exams is also centralized in the sense that some 10 qualified master or Ph.D. students from the physics department grade all exams. Afterwards the exam papers are returned to the respective instructors and so are the corresponding exam scores by email. After the instructors have entered the scores into the GLS/ASIS information system, the students can check their exam results.

As soon as I have the exam papers I will announce a set of dates and times on which students can walk into my office for checking their own exam papers. If these times happen to be inconvenient, then students can make personal appointments. Although it is a principle right of the students to verify the score of their own exams and I offer plenty of opportunities to do so, it is a big surprise to me how few students actually use this right. For example, of the 140 students in the two General Physics 2 courses in Fall 2010 only 13 students checked their midterm and 10 their final exam paper. These figures are representative for all other semesters I have taught general physics at SKKU. Maybe because the physics department announces the solutions of the exam on the website, it is well possible that most students make their own judgement of their exam performance and are then satisfied with the announced score later.

# Assessments

## Diagnostic test

SKKU offers an introductory course to physics as a pre-university education for high school students who are admitted to the university. This course is given at the end of the Winter break to students who find it necessary to level up their basic physics knowledge before they start the first semester. Therefore a diagnostic test as a level check at the start of the General Physics classes I considered not so useful.

A diagnostic test could come in handy if the instructor wants to create groups or teams which are balanced according to the members' level. I have not used this, because it is still not so obvious to me what to do with the test results and what kind of balance is appropriate for the formation of a team. Should all team members have the same level? Or contrary, is it better to force different levels into one team? And what team formation works best for Korean students?

A diagnostic test at certain intervals during the semester can help the students see their own progress and the instructor can use this as a measure for the teaching effectiveness. The downside of regular tests is that it is rather time consuming and for the relatively short semester at SKKU of 14 teaching weeks it would cause too much pressure on the syllabus. Furthermore, the midterm and final exam also have this function, though with serious consequences for the students if their progress is below average.

## Memorizing vs. creative thinking

A few times I have been invited to help with the making of the midterm and final exams. The key design issue for the instructor is then: “What is it you want to examine?”. When selecting appropriate exam problems, the dilemma often rises whether the question should test the students’ ability to memorize facts, or invite students into creative thinking.

The standard rules forbid the use of books and calculators in the exams. Whatever the students need to know to solve the exam questions, they have to know from the top of their heads. Many, if not most students interpret this as the necessity to memorize all facts they have learned in class and from the textbook. This becomes obvious when exam questions are based on sample problems from the textbook, which produce a very high average score among the students. However, when an exam question requires new thinking, beyond the scope the memorizing, the overall performance of the students drops dramatically.

In order for students to develop a sound analytic and scientific mind, creative thinking in science courses is a first requirement and the exams should be able to reflect these objectives. However, more often they do not. In order to break away from this deep-rooted attitude of meaningless memorizing, I started a discussion with the students in my General Physics 2 courses of Fall 2010. In class I asked them to tell me what they want as a midterm or final exam. I got rather unexpected answers. Some students requested multiple choice questions and others wanted problems taken verbatim from the textbook. This means that students hope for exams which are even more leaning towards memorizing. To counterbalance the discussion I introduced a new concept, namely the use of a so-called ‘cheatsheet’ in the exam. Students can write anything on one A4 paper during their exam preparation and they can use the paper in the exam. The sole purpose of it is to stop students from plain memorizing, but instead focus on understanding the concepts while studying. Students could vote in an icampus survey whether they like the cheatsheet or not. It may not be a total surprise that students rejected it; 61% and 55% of the students in the respective courses said not to like the cheatsheet. Although I am personally very much in favor of the cheatsheet, I respected the democratic outcome of the survey.

| Would you like to use a "cheatsheet" in the final exam? | | | | |
|---|---|---|---|---|
| 1. I like the cheatsheet. | 21.21 % | 7 | 41.94 % | 13 |
| 2. I do not like the cheatsheet. | 60.61 % | 20 | 54.84 % | 17 |
| 3. I don't care (no opinion). | 18.18 % | 6 | 3.23 % | 1 |

*A survey in the two General Physics 2 courses of Fall 2010 on whether or not the students like the use of a 'cheatsheet' in the final exam.*

In my opinion the discussion has to be continued on how to modify the exam style in a direction where students have to show a certain level of creativity. This should take place in conjunction with an appropriate teaching style, so that students are properly prepared for what is expected from them in exams. However, this cannot be a single person's grail, but needs to be embedded in a wider discussion together with the responsible people in the physics department.

### Exams

Students should get prepared for the midterm in the 7 weeks prior to the exam. After the midterm another 7 weeks until the semester end prepares the students for the final exam. A few weeks before each exam the physics department publishes all previous exams on the website (http://physics.skku.ac.kr/), so that students get an idea what they can expect.

Students sit for 60 minutes in the standardized midterm and final exams, organized by the physics department. The results of both exams contribute equally to the final grade and in my General Physics courses they can contribute 80 to 100% to the score for the final grade. The midterm and final exams are thus the most dominant happenings concerning the grade. The midterm exam is a critical moment for many students. If the result is below expectation, then it is difficult to overcome the mental blow and regain the motivation to study for a better result in the final exam. I still have not yet found a good solution what to do with the students who are on the brink of giving-up after the midterm exam.

## Grading

For many years SKKU uses grading on the curve (or "norm-referenced grading") as long as the number of students is large enough. Unlike absolute grading, which is entirely performance based, with grading on the curve the grade of a student is relative with respect to the performance of other students in the group. The university uses two conventions for grading on the curve. Standard grading limits the number students with A and A+ to 35%, and the number of students with A, A+, B, and B+ to 65%. If a course is given in English (i.e. "international language") these numbers are raised to 50% and 90%, respectively. The motivation is probably twofold: the prospect of a better grade animates students to take courses in English language who otherwise might not do so, and for certain students the more generous grading policy compensates for possible language difficulties.

Until the year 2009 the General Physics courses were offered in both Korean and English language and students could choose according to their language skills, physics abilities, and grade expectations. In a move towards further globalization, all general physics courses are now taught only in English language, by decree of the SKKU administration head office as of 2009. Although then now the generous “international language” grading policy should apply to all the General Physics courses, the physics department has decided against and instead enforces the standard grading policy. The motivation is that there is no language distinction between courses anymore and thus all students face the same difficulties.

In order to determine the grades of the students, I first calculate the total score of each student. The total score is weighted sum of four assessments:

10% quizzes/attendance score,
10% assignments score,
40% midterm exam score, and
40% final exam score.

The quizzes/attendance score is the result from the classroom problem-solving activities. I do not scrutinize the answer papers, but only give it a bird’s-eye view and judge whether students have tried to provide reasonable answers. The score each time is between zero and one, and is then averaged to obtain the 10% contribution to the grade.

The teaching assistant handles the grading of the weekly assignments and every week I receive by email the results of the assignment teams. At the end of the semester I take the average of all assignment results and calculate the 10% contribution to the grades of each team member.

Finally I determine for every student two final total scores:

1. the score according to the weight percentages of quizzes/attendance, assignments, midterm and final (i.e. 10%, 10%, 40%, and 40%, respectively), and
2. the score according to only a 50-50% average of the midterm and final exam only.

I use the maximum of the two scores for the final grade. Students who perform poorly in midterm and/or final exams can raise their scores with the quizzes/attendance and assignments results. On the other hand, students who have good scores in the midterm and final exams, will not experience negative effects from their quizzes/attendance and assignments results. Excellent students who not need attend class regularly and/or not want to spend much time on assignments, can rely entirely on their midterm and final exams.

I explain these grading rules at the beginning of the semester and I warn students not too easily rely on only midterm and final exam results for their grade, especially students who are weak in physics are guaranteed to fail miserably.

Once the total final score is calculated, the grade on-the-curve is determined by the regulations of the physics department. Also the students are well aware of the grading rules and expect from the instructor to follow the specified guidelines. Therefore I always try to stay as close as possible to the allowable grade percentages, namely A and A+ has to be less than 35%, and A, A+, B, and B+ must be less than 65%.

The university grade regulations are only for the grades in the range from A+ to B. The distribution of the lower grades is a combination of the deliberate choice by the instructor and the university culture. Students who perform poor get a C or C+, whereas students who perform very badly get a D or D+; only students who dropped out of the course or skipped an exam will be given an F.

| Grade | General Physics 1 Class 54 | General Physics 1 Class 59 | General Physics 2 Class 41 | General Physics 2 Class 52 |
|---|---|---|---|---|
| $A^+$ | 9.7 % | 12.5 % | 11.6 % | 11.8 % |
| A | 20.8 % | 19.4 % | 20.3 % | 29.6 % |
| $B^+$ | 16.7 % | 15.3 % | 14.5 % | 16.2 % |
| B | 18.1 % | 18.1 % | 18.8 % | 17.6 % |
| $C^+$ | 13.9 % | 15.3 % | 8.7 % | 10.3 % |
| C | 12.5 % | 15.3 % | 4.3 % | 4.4 % |
| $D^+$ / D | 5.6 % | 0 % | 14.5 % | 10.3 % |
| F | 2.8 % | 4.2 % | 7.2 % | 8.8 % |

*Final grades percentages for General Physics 1 (2010 Spring) and General Physics 2 (2010 Fall) courses.*

Students who are dissatisfied with their grade occasionally beg for a raise. I usually take a firm stance, unless there are obvious reasons as to why the grade accidentally was too low. Sometimes students then ask me to lower their grade to F, which I grant most of the time, because this does not seem to be discouraged by the university. However, students should be warned that they cannot retake the same course indefinitely.

# Teaching Evaluation

## Questionnaires

Icampus offers a survey facility, which I have used for asking the students' opinion on the use of a 'cheatsheet' in the exam (the results of the survey are discussed earlier in this report). Creating a survey in icampus is not a trivial task and without an English manual available, it is a long process of trial and error to set up a survey and make it work. On the other hand, icampus has convenient provisions to immediately view the outcome of the survey.

My survey in icampus for the General Physics 2 courses of Fall 2010 was only a single question and I used very simple English sentences. In the classroom and via an icampus announcement I asked all the students to tell me their opinion. Despite of emphasizing that this survey was important and participation was also in the students' own interest, a little less than half of the students voted in the survey. I have no idea why the other half of the students did not use this opportunity to show their point of view. They don't care? Is the use of an icampus survey too complicated for students? Is a survey in English language scary? Or do students fear their vote in an icampus survey is not confidential? I cannot tell.

A by far more successful outcome has a questionnaire printed on paper and handed out personally to the students during the exam. The midterm or final exam is a perfect timing for such a questionnaire because here usually all students show up. In addition, the questionnaire should be anonymous and have a simple as possible layout in order to produce a high response.

Questionnaire / 설문지 — General Physics 2 December 13th, 2010

(this is anonymous; no need to write your name here / 익명으로 하므로, 이름을 표기할 필요는 없습니다)

**Homework Teams:**

- Do you prefer to make homework in a team or alone? ☐ in a team ☐ alone
- Are you satisfied with your team members? ☐ yes ☐ so-so ☐ no

**Weekly Study Hour (Mondays 7–8pm):**

- How often did you go there? ☐ very often ☐ few times ☐ never
- Was the Study Hour useful? ☐ yes ☐ no ☐ I don't know

**In the classroom:**

- Estimate your class attendance: ☐ ≈ 100 % ☐ ± 50 % ☐ < 25 %
- The teacher's English: ☐ is OK ☐ so-so ☐ too difficult
- The quizzes during class: ☐ I like ☐ so-so ☐ I don't like

**Use the backside of this paper to write your own comments (in Korean or English).**

*Questionnaire handed out to the students in the final exam of the 2010 Fall semester.*

For the final exam of the General Physics 2 course of the Fall 2010 semester I prepared a questionnaire on an A5 paper with seven questions on specific issues related to my course and class management. The answers to each question had a scaling of three levels and the students had to check mark which answer suits best their own situation. Students could finish the questionnaire within seconds. The response was fantastic: of the 126 students who sat in the exam, 125 returned the questionnaire without any invalid returns! I therefore think that the results obtained are reliable and reflect an honest opinion of all students.

| | | |
|---|---|---|
| Prefer doing homework in a team | 65.6 % | Were you satisfied with the team members?<br>yes 62.2 %<br>so-so 31.7 %<br>no 6.1 % |
| Prefer doing homework alone | 33.6 % | Were you satisfied with the team members?<br>yes 14.3 %<br>so-so 40.5 %<br>no 40.5 % |
| Has visited the extra study hour<br>(very often or few times) | 55.2 % | Did you find it useful?<br>yes 30.4 %<br>so-so 26.1 %<br>don't know 43.5 % |
| Never visited the extra study hour | 44 % | |
| Estimate your own attendance | | almost every class 44.8 %<br>about half of the classes 36.0 %<br>less than a quarter 19.2 % |
| What is your opinion about the instructor's English? | | it's OK 79.2 %<br>so-so 14.4 %<br>too difficult 6.4 % |
| Do you like the quizzes in class? | | I like 43.2 %<br>so-so 40.0 %<br>I don't like 16.8 % |

*The questionnaire results of the two General Physics 2 courses in the Fall semester of 2010.*

Analyzing the feedback of 125 questionnaires is unfortunately a laborious task. The data of each paper needs to be entered manually into the computer followed by a careful analysis. The final results are very valuable because of the high response of the questionnaire and the feedback gives the instructor a feel of what parts of the course are well perceived by the students, and what parts need improvements.

Some of the questionnaire results have been discussed already earlier in this report in the appropriate sections and I therefore like here to go only briefly over the results. The majority of students appreciates making homework in teams instead of alone. Quite some students are not satisfied with their team members; maybe in the future I should also ask the students to evaluate their own team members and incorporate this in the assignment evaluation. The sincerity of this questionnaire is demonstrated by the honest answers to the students' own attendance, as this result is in line with my own attendance data. Some students indicate not to like the quizzes as classroom activities; I guess these students do not understand why active participation in class helps them and without a clear grade incentive for classroom participation, these students are not interested. This tells me that improvements and rethinking of the quizzes as classroom activities is necessary so that more students understand the benefits of classroom participation.

## Students' feedback in ASIS

Every time students check their exam results in the ASIS-GLS information system, they first have to evaluate the course and answer a few standard questions about the instructor and the course organisation. The objectives of the questionnaire is to get a rather general impression of how the students perceive the teaching qualities of the instructor.

■문항별 응답인원

| 학수번호 | 분반 | 응답구분 | 문항1 | 문항2 | 문항3 | 문항4 | 문항5 | 문항6 | 문항7 | 문항8 | 문항9 | 문항10 | 계 |
|---|---|---|---|---|---|---|---|---|---|---|---|---|---|
| GEDB008 | 54 | 매우긍정 | 25 | 28 | 19 | 21 | 29 | 20 | 20 | 20 | 19 | 29 | 230 |
| | | 긍정 | 28 | 26 | 24 | 24 | 22 | 25 | 25 | 27 | 25 | 23 | 249 |
| | | 보통 | 16 | 15 | 24 | 22 | 18 | 23 | 23 | 21 | 24 | 17 | 203 |
| | | 부정 | 0 | 0 | 2 | 2 | 0 | 1 | 1 | 1 | 0 | 0 | 7 |
| | | 매우부정 | 0 | 0 | 0 | 0 | 0 | 0 | 0 | 0 | 1 | 0 | 1 |
| | | 무응답 | 1 | 1 | 1 | 1 | 1 | 1 | 1 | 1 | 1 | 1 | 10 |
| GEDB008 | 59 | 매우긍정 | 35 | 38 | 36 | 35 | 35 | 31 | 34 | 35 | 36 | 36 | 351 |
| | | 긍정 | 18 | 16 | 15 | 14 | 17 | 18 | 17 | 17 | 14 | 15 | 161 |
| | | 보통 | 13 | 12 | 14 | 17 | 14 | 15 | 14 | 14 | 15 | 15 | 143 |
| | | 부정 | 0 | 0 | 1 | 0 | 0 | 2 | 0 | 0 | 1 | 0 | 4 |
| | | 매우부정 | 0 | 0 | 0 | 0 | 0 | 0 | 1 | 0 | 0 | 0 | 1 |
| | | 무응답 | 2 | 2 | 2 | 2 | 2 | 2 | 2 | 2 | 2 | 2 | 20 |

■문항별 응답인원

| 학수번호 | 분반 | 응답구분 | 문항1 | 문항2 | 문항3 | 문항4 | 문항5 | 문항6 | 문항7 | 문항8 | 문항9 | 문항10 | 계 |
|---|---|---|---|---|---|---|---|---|---|---|---|---|---|
| GEDB010 | 41 | 매우긍정 | 31 | 30 | 27 | 28 | 31 | 29 | 28 | 28 | 29 | 34 | 295 |
| | | 긍정 | 13 | 15 | 17 | 17 | 18 | 17 | 14 | 14 | 17 | 10 | 152 |
| | | 보통 | 18 | 17 | 16 | 16 | 12 | 16 | 18 | 21 | 15 | 19 | 168 |
| | | 부정 | 0 | 1 | 2 | 2 | 2 | 1 | 3 | 0 | 2 | 0 | 13 |
| | | 매우부정 | 2 | 1 | 2 | 1 | 1 | 1 | 1 | 1 | 1 | 1 | 12 |
| | | 무응답 | 1 | 1 | 1 | 1 | 1 | 1 | 1 | 1 | 1 | 1 | 10 |
| GEDB010 | 52 | 매우긍정 | 38 | 39 | 35 | 36 | 40 | 39 | 33 | 35 | 35 | 39 | 369 |
| | | 긍정 | 19 | 17 | 19 | 19 | 16 | 15 | 17 | 21 | 15 | 16 | 174 |
| | | 보통 | 8 | 9 | 12 | 9 | 9 | 13 | 14 | 8 | 13 | 12 | 107 |
| | | 부정 | 2 | 2 | 1 | 3 | 2 | 0 | 2 | 3 | 4 | 0 | 19 |
| | | 매우부정 | 0 | 0 | 0 | 0 | 0 | 0 | 1 | 0 | 0 | 0 | 1 |
| | | 무응답 | 1 | 1 | 1 | 1 | 1 | 1 | 1 | 1 | 1 | 1 | 10 |

*The evaluation of the objective questions of the students' feedback in the ASIS-GLS information system for the two General Physics 1 courses (upper panel) and for the two General Phycsis 2 courses (lower panel) in the Spring and Fall semesters of 2010, respectively.*

At first there is an objective evaluation, where students must answer questions numerically on the scale from low to high. Between 70 and 80% of all students have selected the highest or next to highest scale. This is a very flattering outcome for the instructor, especially when taking into account that there are always some 10% of the students in a course with minimal interest in physics.

The objective evaluation is followed by a set of subjective questions, which invites students to write their personal response. Most students answer in Korean, but a few make the effort to write in English and here are some of their responses:

- *“This class was worthwhile. Thank you for teaching me during this semester.”*
- *“I hate team class.”*
- *“Professor's explanation is very difficult.”*
- *“I'm very satisfied with the good lectures. I hope to see you again in another course.”*
- *“This lecture is very good.”*
- *“GOOD”*
- *“Solving textbook exercises might be another good way to improve understanding.”*
- *“It's too difficult.”*
- *“Homework is too much.”*
- *“Very great class! With this lecture, I became interested in physics.”*
- *“Team work is a good idea.”*
- *“Homework is so difficult.”*
- *“I attended little, but I watched the recordings on icampus; this class was excellent.”*

The questions the students are asked in the questionnaire are in English (as far as I am informed). Some of the questions are quite confusing and difficult to unravel for students who are non-native speakers of English. For example:

> *“There were no class cancellations due to the instructor's personal matters. If classes had to be cancelled due to unavoidable reasons, the instructor held sufficient make-up classes.”*

The answer to this question is ambiguous; if a student says “Yes” or is “Satisfied”, then does this mean that there were no cancellations? Or sufficient make-up classes were held? And what if a student says “No” here?

The outcome of the ASIS-GLS course evaluation can give an overall feel for how well or badly the students judge the instructor’s teaching style. However, in detail the evaluation should be taken with a pinch of salt, because the nature of the questions is general and at times even ambiguous.

## Peer observation reports

A peer report is a teaching evaluation performed by a colleague instructor. This is usually an incidental class observation where both the instructor and observer agree on the time and date. A peer report has the interesting aspect that it can be helpful to the instructor as well as the observer. Both the instructor and the observer need to be well aware of the facets of different teaching strategies and how objectives can be achieved through an appropriate teaching method. Therefore a good observer is as valuable as a good instructor!

In the appendix is an example form of a peer report for one General Physics class with Natanael Karjanto as the observer. The instructor has to state the objectives of the class and mention particular issues about the class under observation. The observer uses this information to evaluate the teaching style.

I suggest to seek for a possibility to implement a peer report system throughout the university where instructors are requested to become observers of their colleagues.

# Calculus Course

## In the classroom

### Course management

Two Calculus courses (Calculus 1 and Calculus 2) are by far the largest classes in terms of student's number registered at SKKU. During Fall 2010 semester, we have almost 1600 students taking Calculus 2 courses and the majority of them are freshmen students. There are 12 instructors handling a total of 18 classes and one instructor acts as a course coordinator. The maximum number of students in one class is 80 pupils and there are slight variations in the number of students from one class to the other. All students sit for identical examination papers, both for midterm and final exams. However, each instructor may have additional assessments, such as assignments, group projects and/or quizzes.

In addition, during Winter 2010 semester, the Department of Mathematics offers two classes in Calculus 2 and both are conducted independently by two different instructors. As a consequence, all exams are also conducted independently. Since the Winter semester is short and intensive, there are neither graded assignments nor group projects. However, there are a few voluntary and ungraded assignments and group projects to gauge the students' interest in learning. Similar to the regular semester, the maximum number of students in one class is also 80 pupils.

### Textbook and syllabus

We adopt the same textbook for the Calculus 1 and Calculus 2 courses, both in regular (Spring and Fall) and in short semesters (Winter and Summer). Although there is a plan to use a different textbook for Calculus starting the new academic year 2011, so far we have been using the book authored by James Stewart: Calculus-Concepts and Contexts, Metric International Edition, published by Brooks/Cole Cengage Learning, 2010. The Korean translation of this book is also readily available to the students and some of them prefer the translated version instead of the original one for a better understanding of the material. As already addressed above for the case of the Physics textbook, a similar situation occurs with regard to Calculus textbook. It is very sad to observe that many students made photocopies of the book instead of purchasing the original version of it. We strongly urge the university to do something about this violating of copyright law.

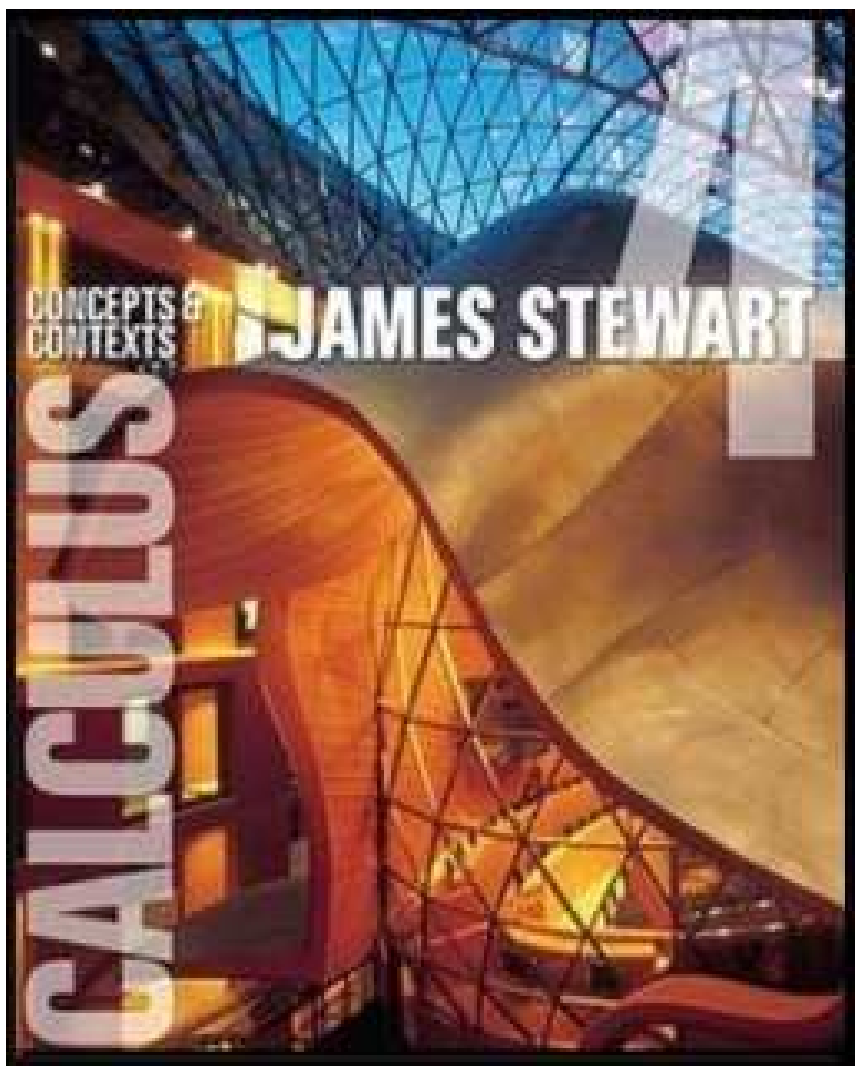

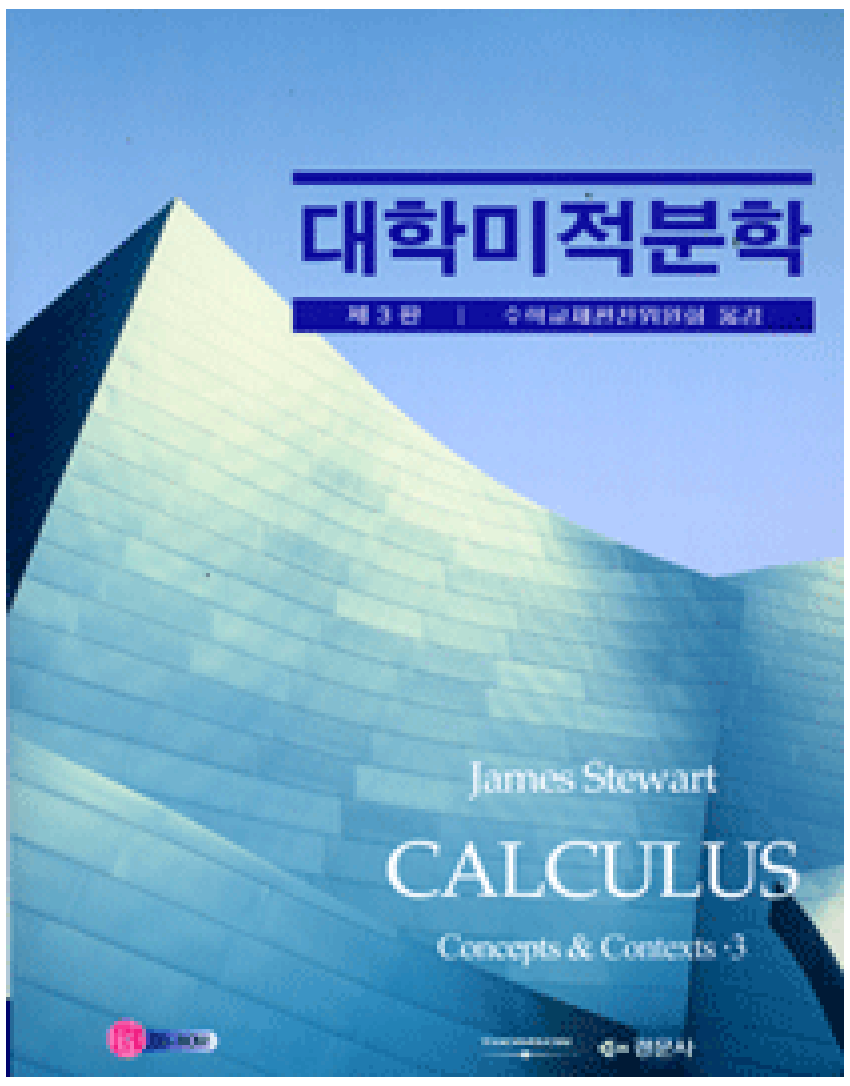

*The cover of Calculus textbooks both for the English and the Korean versions.*

This textbook from Stewart is intended for a year long course, covering both single and multivariable Calculus, including an introduction to Vector Calculus. To obtain a general overview of the syllabus for the two Calculus courses, we may simply divide the textbook into two parts. Calculus 1 covers from Chapter 1 to Chapter 8 and Calculus 2 covers Chapter 9 to Chapter 13 of the book. Although more chapters are weighted to Calculus 1 course, the pace of teaching can be done faster since the students are knowledgeable already with many parts of Calculus from their high school period. On the other hand, even though only five chapters are covered in Calculus 2, there are occasions that we need to rush with the material and skip some parts in order to explain better on other parts, which are rather new to the students. In fact, the course coordinator has provided a detailed syllabus, including the parts that need to be skipped and to be omitted. The students may access the syllabus through icampus system of the university as well as the ASIS-GLS information system. The semester at SKKU runs for 16 weeks, including midterm and final term exams. So, basically we have 14 weeks of effective teaching and each week is composed as two one-and-half hour session or one three-hour session, depending on the timetable provided by the Mathematics Department.

The textbook from Stewart is an excellent material for students to learn Calculus, as many of us who are teaching have observed and agreed upon. It presents a balance between theoretical concepts, worked examples, use of technology, short biography of some mathematicians and what is very important for the students is its feature in providing many exercise problems with diverse level of difficulties. The book is intended to be a mainstream Calculus text that is suitable for every kind of course at every level. It is designed particularly for the standard course of two semesters for students majoring in Natural Sciences and Engineering. Certainly, students are expected to have some background of high school mathematics in algebra, geometry and trigonometry. The book also contains many beautiful artworks to give a vivid illustration to readers when certain concepts are being explained. For every edition published, the author always makes an endeavour in improving the textbook, this includes the mathematical precision, accuracy, clarity of exposition and outstanding examples and problem sets that have characterized the earlier editions. The focus on problem solving has made the book a favourite of students and instructors in a wide variety of colleges and universities throughout the world. The applications of Calculus in many fields of study are also presented in an interesting manner. Since the book is rather challenging, the students who struggle with mathematics may have a difficult time with the book's rigorous and some algebraic or conceptual jumps.

The first three editions of the textbook were intended to be a synthesis of calculus reform and traditional approaches to calculus instruction. In the fourth edition, the author continues to follow that path by emphasizing conceptual understanding through visual, verbal, numerical and algebraic approaches. The aim is to convey to the students both the practical power of Calculus and the intrinsic beauty of the subject. In whatever teaching approaches that one instructor may implement, or in whatever style approach the book may present, a common goal is remained to be achieved: to enable students to understand and appreciate Calculus. Some features of the book consists conceptual exercises, graded exercise sets, real-world data, applied projects, laboratory projects, writing projects, discovery projects, problem solving, some rigour proofs, use of technology, Tools for Enriching™ Calculus and Enhanced WebAssign.

## Implemented teaching method

The teaching method implemented for Calculus courses, and any other mathematics based courses, is generally traditional and conventional: the instructor acts as the controller of the learning environment. (Visit the website of Flinders university on education.) The instructor of traditional way creates an environment for which the learning process is teacher-centered instead of student-centered (Novak, 1998). Thus, in a number of occasions, the teaching session is heavily dominated by delivering a lecture to the students with a small or even no participation from the students' side. Nowadays, many expertise in education around the world encourage many educators, including school teachers and university instructors to combine, if not switching entirely, to problem-based learning and more students' active participation. There are definitely abundant literature on these topics, but amongst others are discussed by Rosenthal (1995), Röj-Lindberg (2001), Zhang (2002), Ramsay & Sorrell (2006) and Zwek (2006).

The fact that many suggested teaching methods move away from the conventional one, does not mean that this traditional teaching method is not useful anymore. Depending on how we implement the traditional teaching method, an effective teaching that creates a conducive learning environment and gives an optimal benefit to the students can be achieved. Good preparation, structured organization, generating an outline, analyzing the audience, choosing examples, choosing learning activities, using audio/visual aids and reviewing the materials are some possible ways to make an interesting lecture with a traditional teaching method.

For further information, visit the website of Center for Teaching Excellence at University of Medicine and Dentistry of New Jersey. Although some might believe that non-conventional or modern teaching methods are better than the traditional and conventional ones, alternative methods are equally successful when they are handled in an effective manner.

In Calculus courses, entirely abandoning the traditional and conventional teaching method is impossible. However, we could include elements of non-conventional teaching methods so that teaching activities will not only have more variations but also become more interesting for the students. In turn, it is our expectation that they will have a voracious appetite in learning and studying. Similar to many textbooks for basic courses at the freshman level of undergraduate programmes, our Calculus textbook provides excellent teaching materials. There are ancillaries for both the instructors and the students. For instructors, the PowerLecture CD-ROM is provided. This includes lectures in Power Point format and an electronic version of the Instructor's Guide. Apart from this, the instructor also has an access to the accompanying Complete Solutions Manual of the textbook and Tools for Enriching™ Calculus available at www.stewartcalculus.com. Ancillaries for students include Stewart Specialty Website (www.stewartcalculus.com), Enhanced WebAssign, The Brooks/Cole Mathematics Resource Center Website (www.cengage.com/math), Maple CD-ROM, Tools for Enriching™ Calculus, Study Guide for both Single and Multivariable Calculus, Student Solutions Manual and CalcLabs with Maple and Mathematica.

The PowerLecture CD-ROM is fully utilized during the teaching activities. When explaining new concepts, definitions, properties and theorems, the CD-ROM is very useful since it saves time in comparison to re-writing again everything on the blackboard/whiteboard. The CD-ROM also contains good quality artworks that describe mathematical concepts or three-dimensional objects which are difficult or even impossible to draw by hand. When discussing examples, the CD-ROM provides a step by step explanation, identical to the textbook. Here we can easily lose the attention of the students. So, instead of directly displaying the step by step explanation from the slide, it is good to explain the students by hand, either by writing on the screen or on the whiteboard. By doing this, hopefully we are able to keep the students' attention and improve their understanding in problem solving. After this step, we could ask the students whether there are unclear steps or whether they have questions. Usually, they keep quiet and have no questions though. Then we could proceed to following topic or assign them exercise problems from the textbook. These exercises maintain high alertness of the students.

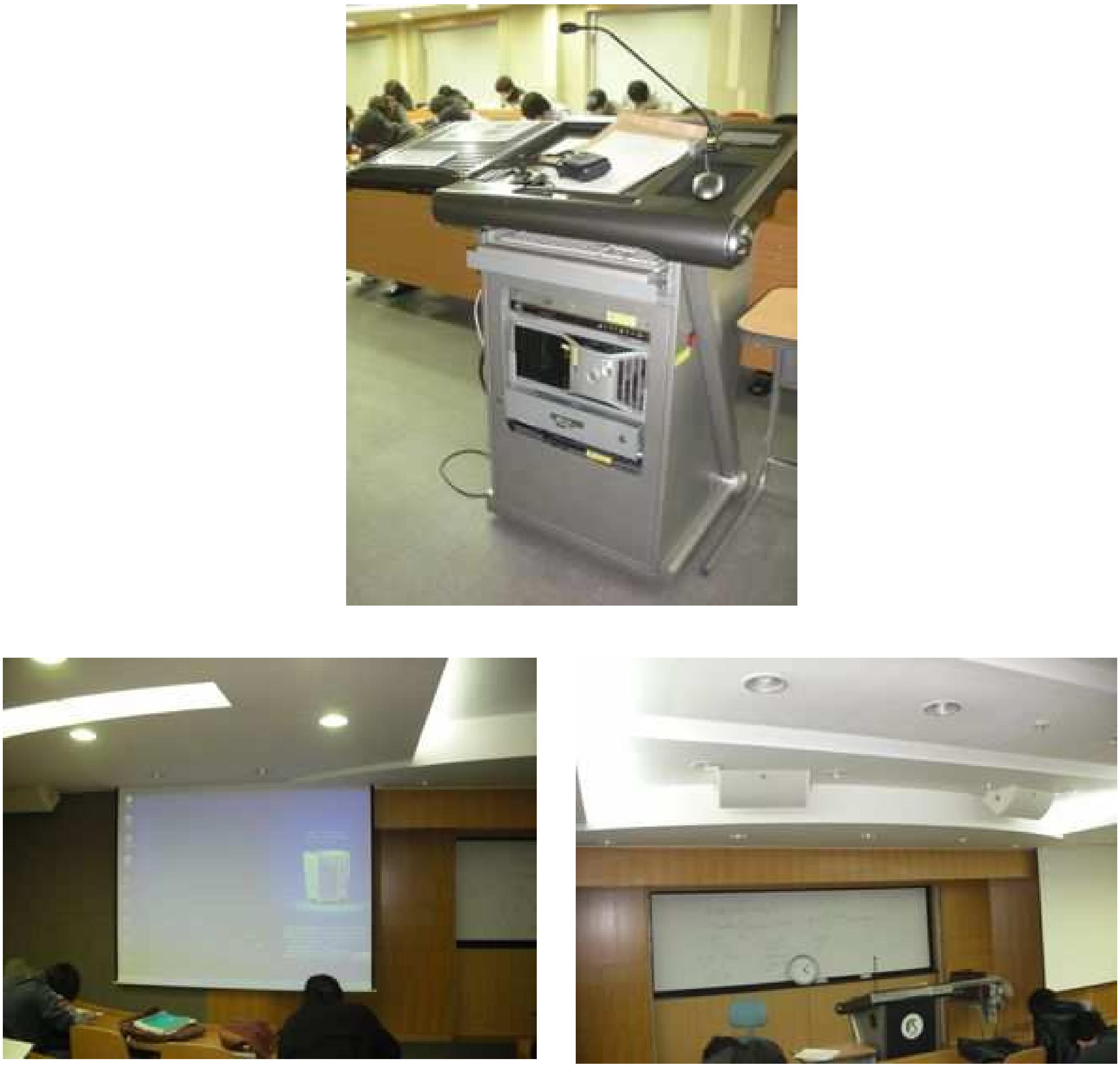

*Pictures taken in the classroom od the Calculus class. Each classroom is equipped with whiteboard and a computer connected to a projector.*

Every lecture session would be recorded (the so-called e+ lecture), and the recording is uploaded automatically to the icampus system after the teaching session is completed. This recording does not only capture the explanation from the instructor but also includes the slides and everything we write on the screen. This is beneficial to the students since they might want to watch the lecture again in case they find certain concepts difficult to grasp and the pace of instructor's explanation is too fast but they are too shy to ask questions or to ask the instructor to repeat again, particularly certain difficult concepts. The recording is also useful for the students who miss the class during that particular session of the day.

The classrooms used for Calculus 2 classes during Fall and Winter 2010 are 31255 and 26515, respectively. Both classrooms are lecture hall type and the desk and chairs are fixed. These classrooms are lecture hall type and re-arranging seat is not possible. Since the students always sit at the back of the classroom or at the far side seats, I always encourage them to take the front and the middle seats. Of course we cannot force them, but it is preferable that as many students as possible use these seats and leave the rest for those who come late. The response to this admonition is rather poor. Although some students move to the front seats, the majority still remain at the back of the classroom. Very often, the first two or three rows of the front seats are empty for room 31255 and the first front row is empty for room 26515. Furthermore, the students prefer to sit close to their peers whom they know. When the students sign up a particular class, they may have some discussions and agreements with their peers when choosing a certain lecturer. This may explains why some students prefer to mingle with their peers rather than mix around with other `non-familiars'. On the other hand, as the semester evolves, some students make new friends with others whom initially they do not really know. Signing up the class is on the first come first served based. As a consequence, the composition of one class may vary not only on their major study but also in their level year.

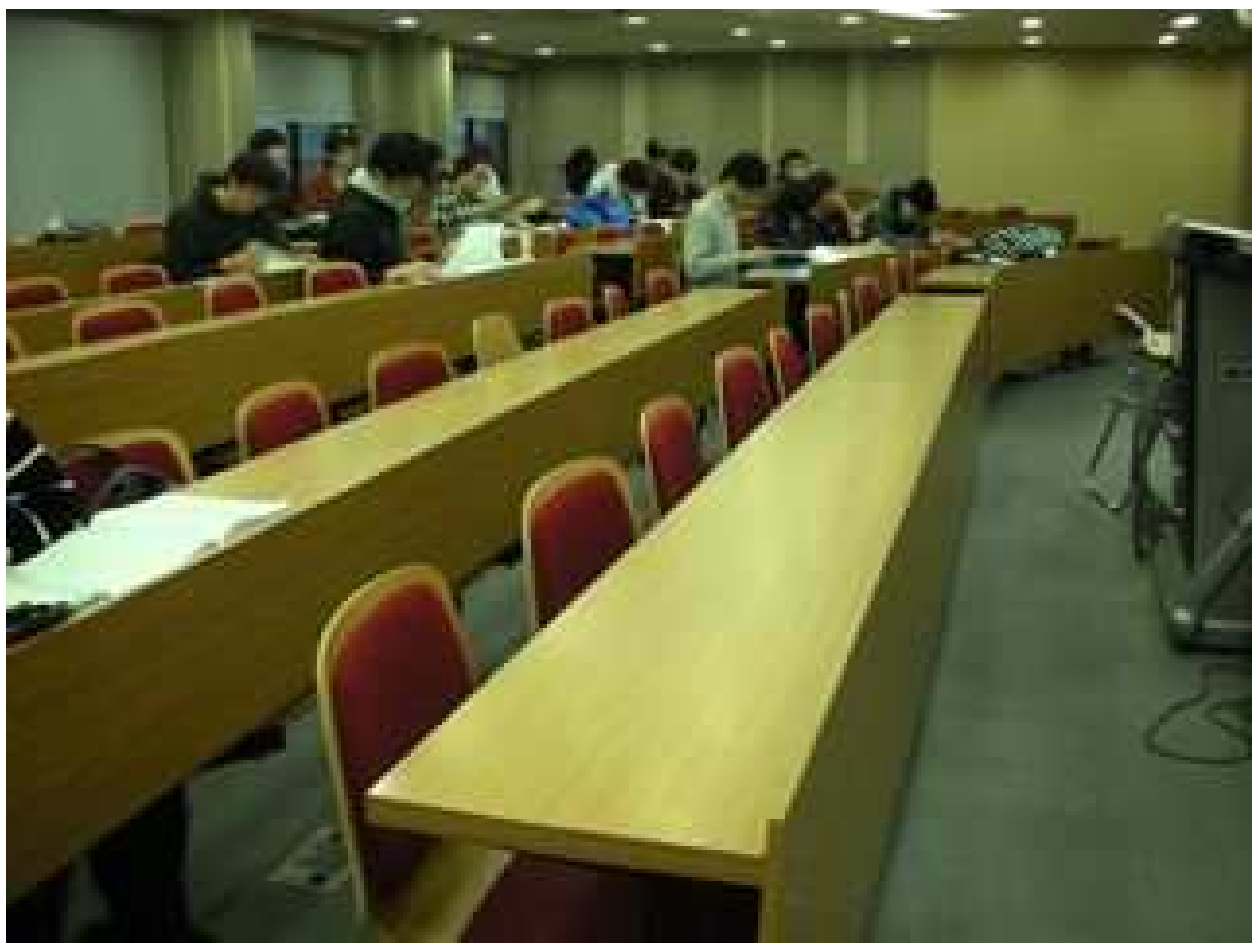

*It is not uncommon that the front row of the classroom remains unoccupied for the entire semester.*

## Attendance record and students' participation

The attendance of the students during the class session is recorded either by calling their names or just by distributing the attendance list and let the students tick their own names. However, both techniques present their own challenges. Calling students' name one by one for every teaching session can be time consuming and with the risk of not being able to cover the entire material. Yet, this is by far the most accurate record for students' attendance. On the other hand, by letting the students tick their own attendance list presents loopholes in term of dishonesty. Those who intend not to attend the class might ask their peers to tick the attendance list for them. This method is time effective, provided that the students distribute the attendance list to those who have not ticked yet. However, from my experience, it is common that the students do not pass it to the others and just let it lie on the desk.

The students' active participation can be stimulated by posing them several questions during the teaching session and allow them to give a response. When I solve an example from the textbook, it may happen that I make some calculation mistakes. When this situation occurs, we could ask the students to pinpoint where the errors are. It helps them to think and stay involved with the teaching session. Another way is by asking one or more students to write and present their work on the whiteboard. From here, I could monitor the way they present mathematical expressions and this also reflects the way they write their answer in the exam. From this I could observe that we can improve the way the students present their work. Even if it is no problem to find the answer to the question from the textbook, the presentation of the answer may need to be improved. Furthermore, when I asked them to explain their answer, many of the students find it difficult to explain in English although they can explain easily in Korean. So, after the students complete their work on the whiteboard, we could give further training by asking them some questions of what they have written and let them explain with their own words. It is very often discovered that they write certain symbols but did not explain what those symbols mean. This training is hopefully useful for them when eventually they have to write their exam papers.

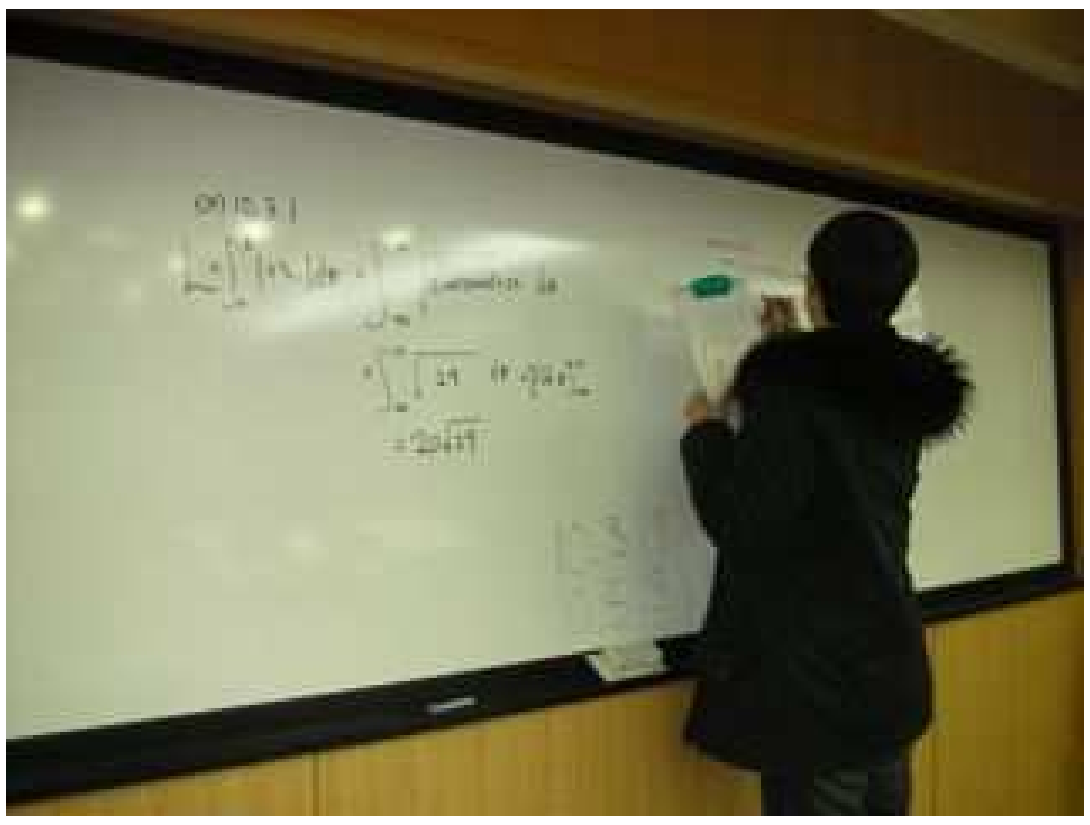

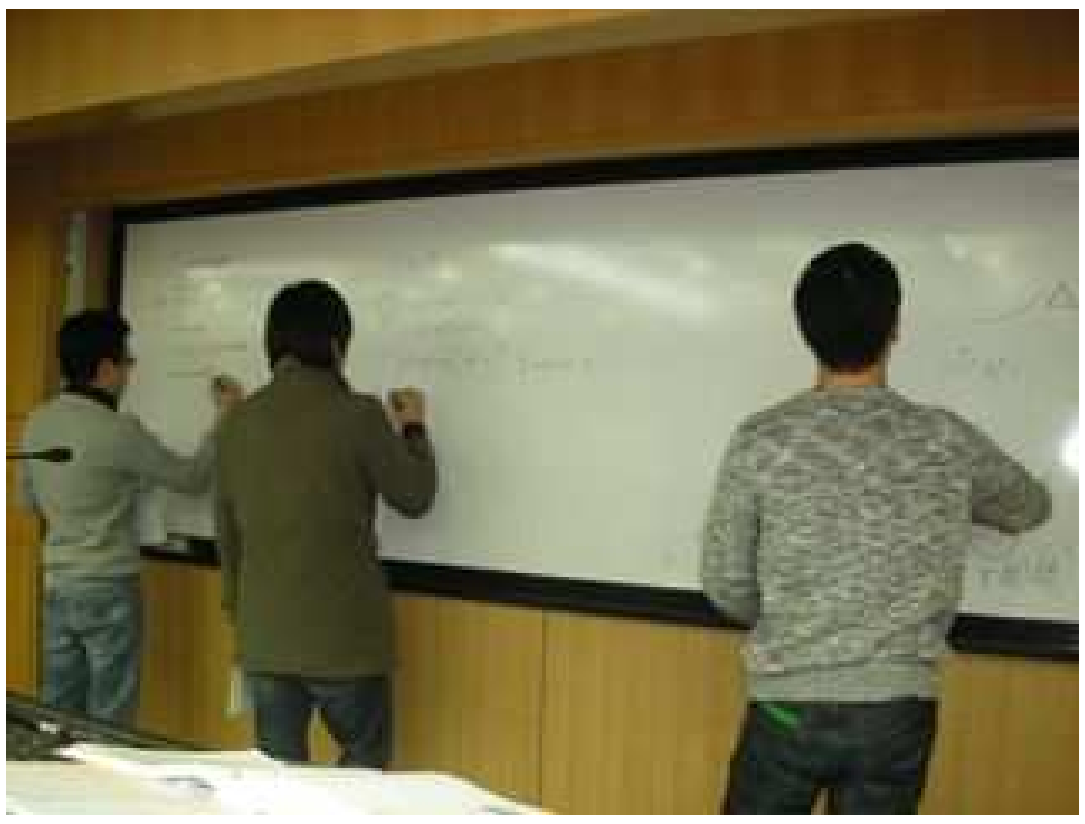

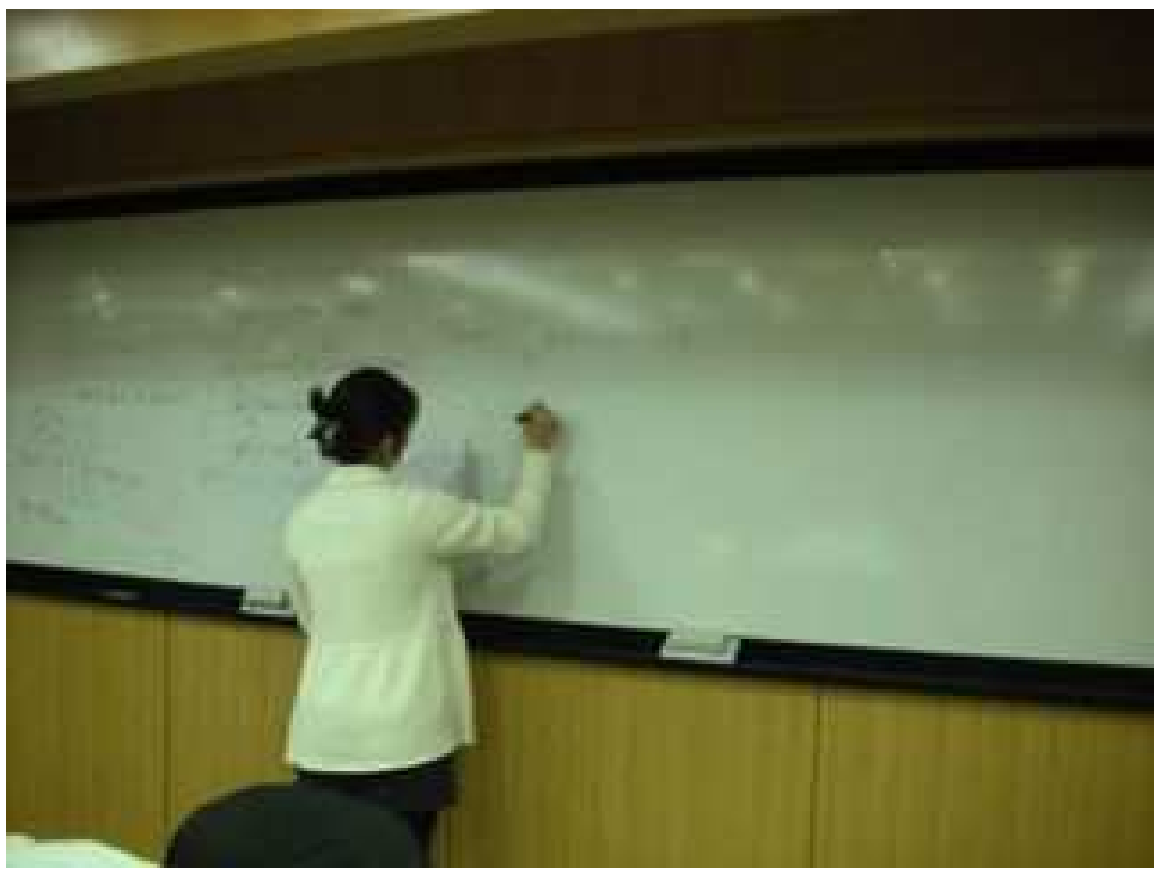

*Photographs of students when they are asked to write their solutions on the whiteboard of the assigned exercises during the class session.*

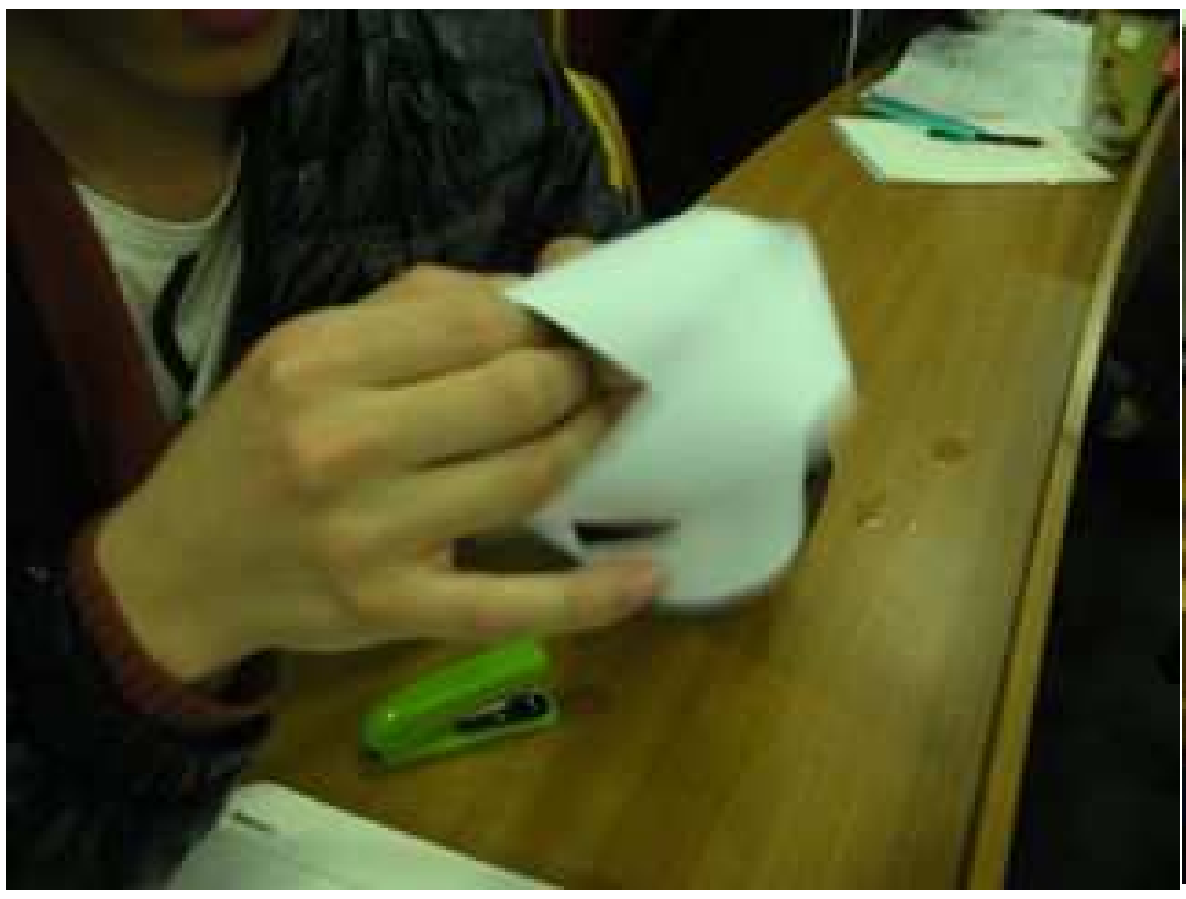
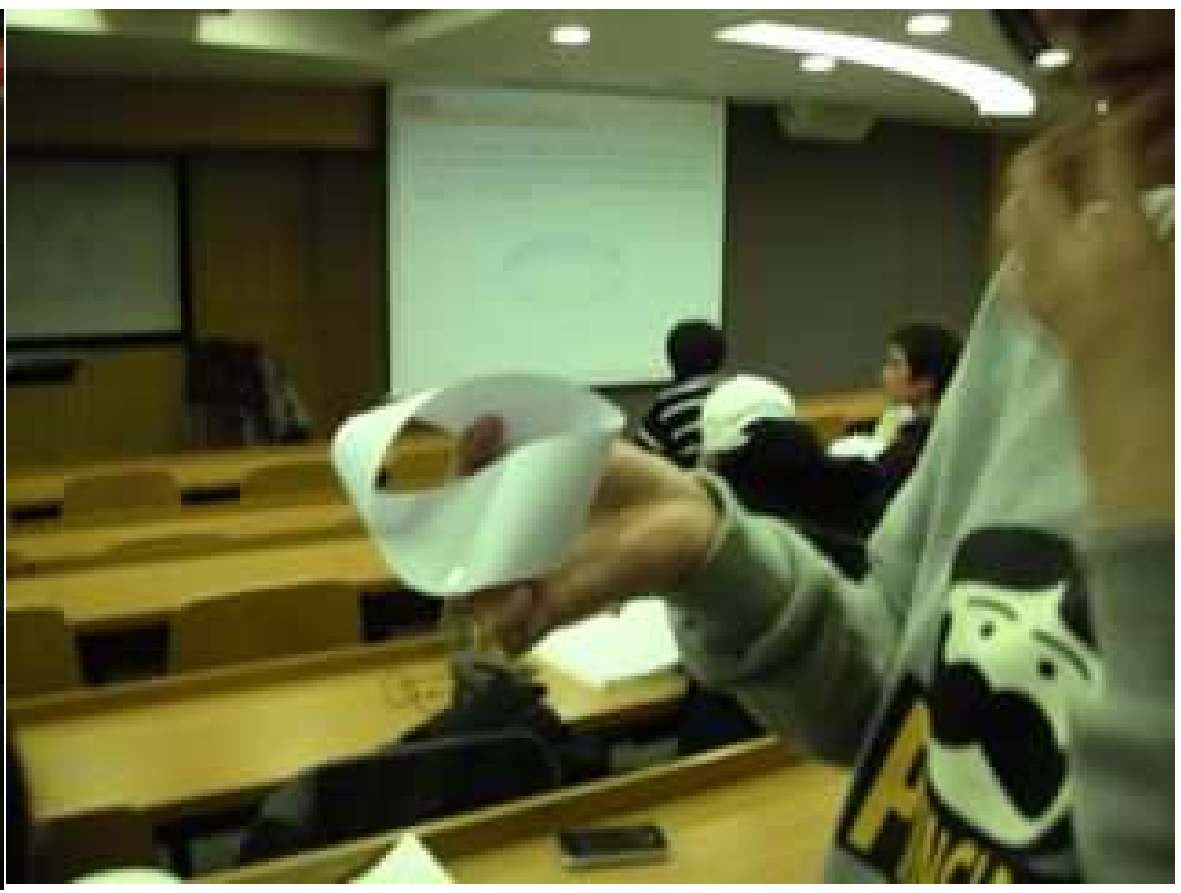
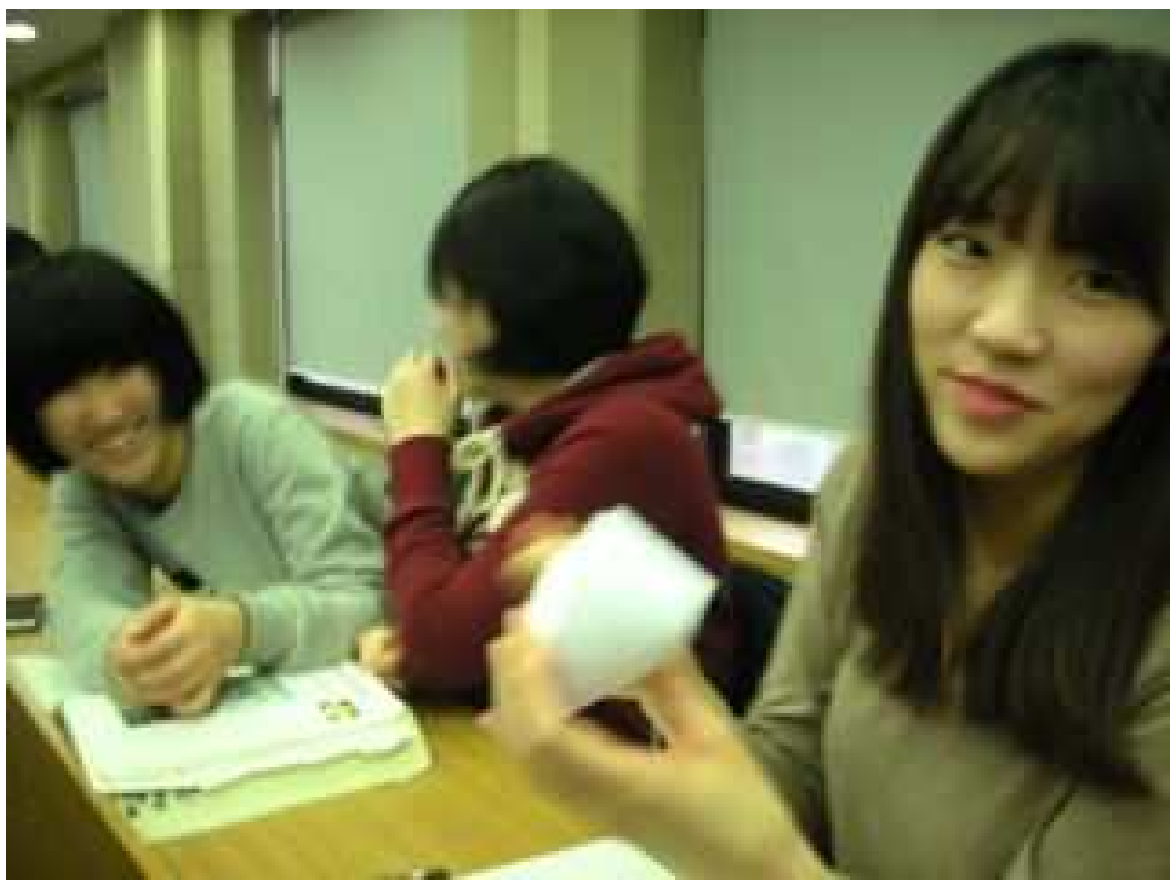

*More photographs of class activities. The students are asked to construct the Möbius strip using paper and tape. The students are happy and enjoy such fun activities.*

# Out-of-classroom

## Assignments

During Fall 2010 semester, there are four individual assignments that the students need to submit as part of their assessment to receive a final grade. These assignments are taken from the textbook and each set of assignments contains 20 questions. In the textbook, only the question numbers printed in red colour are assigned as homework. There is no specific group assignments during this semester but the students have the freedom to work in a group if they wish and submit their assignments individually. The assignments are distributed around 10-14 days before the submission deadline. Late submission will not be accepted. Two assignments are given before the midterm exam and the other two are given before the final term exam. The students submit their assignment in a designated pigeon hole in a seminar room in the Department of Mathematics. The TA then collected their assignments, graded them, gave remarks and returned them to the same room, where the students may collect their assignments. The weight of the assignment is 10% of the final grade. The majority of the students are very diligent in handling in their assignments. Only a few number of students did not submit all the assignments whereas some of them missed one or two assignments out of four.

During Winter 2010 semester, the homework assigned to the students is not graded and thus did not affect their final grade. Submission to these assignments is on a voluntary basis: those who are willing to learn more and do the exercises from the textbook are encouraged to hand in their reports. The response is good, around 80% of the students submit their homework despite the fact that it is assigned only one or two days earlier. (Of course, the number of questions is also kept to be minimal, around one or two questions each.) Since the nature of this assignment is voluntarily, the students are allowed to submit either individually or as a group. It is interesting that more than half of the students prefer to submit the assignments individually rather than as a group. Furthermore, recommended groups have been formed based on a diagnostic test conducted at the beginning of the semester. Each group consists of three students, so there are 25 groups designated as Group A, B, C …., Y. However, not a single student has followed this recommended groups since they prefer to mingle with their peers, the friends that they know earlier instead of mingling with people they do not really know. So, when the students submit the assignments as a group, they choose to to work with their own self-formed groups rather than the recommended groups. We will discuss further on this diagnostic test on the section of Assessment and Examination.

## Contact hour with the teaching assistants

The students who enrolled in Basic Science and Mathematics courses are given an additional privilege to get help from the TAs with the provision of contact hour for an hour each week. During this contact hour, the students have an opportunity to ask questions either on theoretical and concept parts or on tips on problem solving. We are convinced that those who make full use of this provision will benefit greatly and could boost their confidence for the exams. The time for Calculus contact hours is every Wednesday evening from 6 pm until 7 pm, when the students do not have an obligation anymore to attend any other lectures. As instructors and professor, we have been asked specifically to give an announcement to the students and to encourage them to attend this contact hour session. However, we receive reports from our TAs that the attendance for this provision is very poor. On average, every session is attended by none or one student. Perhaps the university or the university college could think of an alternative way to improve the attendance for this contact hour session, in particular for the students who are rather weak in Mathematics, Physics and Chemistry.

## Office hours

An office hour has been provided to the students so that they could approach the instructor outside the teaching session if they feel too shy to approach during the teaching session or during the break. For Mathematics courses (including Calculus), the office hour is every Monday and Thursday at 1:30 pm - 3:00 pm. A similar situation to the contact hour with the TA, the office hour is even worse, with zero attendance during Fall and Winter 2010 semesters. Nevertheless, a number of students ask their questions after the teaching session ends. This may give some level of satisfaction for them so that they might think it is not necessary for them to come again to meet the instructor during office hour.

In order to stimulate the students' interest in making full use of office hour, we could remind them again and again regarding this important provision. We also could make them curious by posting a challenging problem in the classroom, let them think about it, arouse their curiosity and invite the students to come for the office hour if they would like to know how to solve that problem. Again, this method has a weak point. Those who like to know more are normally good students and those who are weak will lose their interest and simply avoid both the subject and the instructor, just hoping to pass and get away the subject as soon as possible. If this is the case, there is no special remedy but simply trying our best as an instructor.

# Assessments

## Diagnostic test

A diagnostic test in Mathematics is a common practice throughout many universities in Europe and North America. For examples, the reader may visit the websites of School of Mathematics, The University of Manchester, UK and Department of Mathematics at University of Calgary, Canada. A diagnostic test may be used to waive certain prerequisite requirements to follow a certain course provided that the students achieve satisfactorily grade in this test. Normally, the test is designed to access general problem solving skills from elementary or secondary school mathematics in areas such as arithmetic, algebra, geometry, trigonometry and logical thinking.

There are a number of reasons as to why diagnostic tests are necessary. According to Dr. Steele from School of Mathematics, The University of Manchester, a diagnostic test is conducted in order to give individual students an idea of their strengths and weaknesses, to build up profile of class as a whole, for instance in tailoring the courses and to identity the students most in need of extra help. Furthermore, there are some supporting literature and evidence of successful results regarding some implementations of diagnostic tests in Mathematics. For instance, Edwards (1997) addressed that the diagnostic test combined with follow-up support can be successful to the students who are initially weak in Mathematics. From a different perspective, Halloun and Hestenes (1985) examined the initial knowledge of college Physics students with diagnostic test in Mathematics, which includes five Calculus items omitted in College Physics. Furthermore, Meltzer (2002) investigated that there is a strong correlation between the students' ability in mathematics for a success performance in college Physics courses. In addition, Robinson and Croft (2003) described some recent strategies adopted to support first year engineering students through diagnostic test and its follow up as well as the success upon these strategies. Haßler, Atkinson, Barry and Quinney (2003) presented the results of a survey of university freshmen students into the experience of mathematics diagnostic testing. Although a large majority of the students feel that such tests were beneficial, many also feel that they were not prepared.

For Calculus, a similar diagnostic test can be conducted at the beginning of the class session. The Calculus textbook by James Stewart provides diagnostic tests on four different topics: Basic Algebra, Analytic Geometry, Functions and Trigonometry. These tests are given at the beginning of the book and answers are given so that the students who do not do well may discover their own mistakes. The authors stated that success in Calculus depends to a large extent on knowledge of the mathematics that precedes Calculus, i.e. the topics mentioned above. The provided tests are intended to diagnose weaknesses that the students might have in those areas. The authors also encourage the students to refresh their skills if necessary by referring to the materials provided.

MATHEMATICS DIAGNOSTIC TEST

**Problems:**

1. If there are 3 apples and you take away 2, how many do you have?
2. You are attending a meeting and during that meeting there are 50 people present. If you shake hand with all the attendants, how many times you shake hand?
3. Two boys and two girls shake hand among each other. How many times they shake hand?
4. Toni is thinking of a number. If you double the number and add 11, the result is 39. What number is Toni thinking of?
5. If a brick weights one kg plus half of its own weight, how much does it weigh?
6. The number of chickens and rabbits are 4. The number of their legs are 10. How many chickens and rabbits each?
7. Larry has exam scores of 59, 77, 48 and 67. What score does he need on the next exam to bring his average for all five of the exams to 70?
8. Two years ago it took Tom eight hours to whitewash the fence. Last year, Huck took just six hours to whitewash the fence. This year, Tom and Huck have decided to work together, so they'll have time left in the afternoon. How long will the job take the two boys?
9. Enrique and Ana each ordered a large pizza. Enrique likes large pieces, so he requested his pizza be cut into 8 pieces. Ana, however, requested that her pizza be cut into 12 pieces, since she likes smaller pieces. Enrique ate 6 pieces of his pizza and Ana ate 9 pieces. Who eat more pizza, Enrique or Ana?
10. The new city park will have a $2\frac{1}{2}$ acre grass playfield. Grass seed can be purchased in large bags, each sufficient to seed $\frac{3}{4}$ of an acre. How many bags are needed?
11. On a cruise to the Caribbean, 2/3 of the women passengers on board are married to 5/6 of the men passengers. There are also 9 more women than men on board. How many passengers are on the ship?
12. Amy paid RM 36 for a pair of shoes at the 'one-fourth off' sale. What was the price of the shoes before the sale?
13. Khalid and his brother Ahmed both left their home city at 1 pm, but Khalid averaged 63 km/hour in his new car and Ahmed averaged just 51 km/hour in his older car. How much farther has Khalid traveled than Ahmed at 4:30 pm that afternoon?

1

14. Hannah and her brother John went to a shopping mall with an equal amount of money to buy their mother a present. Hannah purchased a RM 26 scarf and John bought his mother a jewelry box for RM 34. After their purchase, John has 2/3 of the money that Hannah has remaining. How much money do Hannah and John now have, and how much did they start with before making their purchases?
15. The pyramid of Khufu is 147 m high and its square base is 231 m on each side. What is the volume of the pyramid?
16. An ice cream cone is 5 cm high and has an opening 3 cm in diameter. If filled with ice cream and given a hemispherical top, how much ice cream is there?
17. A man distributes some apples. On his way out he met three people one after the other. To each one he gives half of the apples he had. In the end, he had only one apple left. How many apples did he originally have?
18. If it were two hours later, it would be half as long until midnight as it would be if it were an hour later. What time is it now?
19. A toy maker has a rectangular block of wood 30 cm × 10 cm × 14 cm. He wants to cut as many 3 cm cubes as possible. How many such cubes can he cut?
20. Peter, James and Ruth had some stamps. James and Ruth together had 3 times as many stamps as Peter. The ratio of the number of stamps James had to the number of stamps Ruth had was 3 : 7. Peter and Ruth had 310 stamps altogether. How many stamps did Peter have?
21. Mr Lau planted 9 seedlings in a row. The seedlings were planted at the same distance apart. The distance between the first and fourth seedlings was 12 cm. What was the distance between the first and ninth seedlings?
22. At first Sara had 4/7 of the number of marbles Jack had. When Sara received 36 marbles from Jack, both had the same number of marbles. How many more marbles did Jack have than Sara at first?
23. The day before yesterday I was 25 and the next year I will be 28. This is true only one day in a year. What day is my birthday?
24. Find the mistake in these mathematical equations.

$$\begin{aligned} x &= 2 \\ x(x-1) &= 2(x-1) \\ x^2 - x &= 2x - 2 \\ x^2 - 2x &= x - 2 \\ x(x-2) &= x - 2 \\ x &= 1 \end{aligned}$$

25. One day Kerry celebrated her birthday. Two days later her older twin brother, Terry, celebrated his birthday. How come?

2

*A sample of diagnostic test given to the students at the first session of the Winter 2010 semester.*

At the very first day of the class during Winter 2010 semester, a diagnostic test has been conducted. Most of the questions test algebraic capability, the logical thinking and the ability to formulate story problems into mathematical formulation. Many of the questions are at the level of elementary and middle school mathematics. Yet, some of them are quite tricky and challenging, require non-routine procedure to solve and definitely force the students to think deeply. There are 25 questions and the students are required to finish within 60 minutes. The final hour of the class session is allocated for this diagnostic test. Out of 78 registered students, only 56 participated in the test. The highest score is 84%, the lowest score is 28%, the average score is 58.25% and the standard deviation is 12.4%. The sample of diagnostic test is given in the figure above.

## Memorization vs. creative thinking

As mentioned earlier, the assessment for Calculus courses is given a high percentage on both the midterm and the final term exams, although a small percentage may be allocated for assignments, quizzes, group projects and other relevant assessment components. For Fall 2010 semester, the midterm and final term exams carry 40% weight each. The other 20% is obtained from four sets of assignments (10%) and two quizzes (10%). For Winter 2010 semester, since the time is limited and the lecture is delivered in a rather intensive manner, the assessment for the final grade is obtained only from the midterm and final term exams, for which each exam carries 40% and 60% weights, respectively. The assigned homework mentioned earlier is submitted on voluntary basis. It is ungraded and would not affect the final grade of the students. The result of the diagnostic test does not affect either.

When designing questions for an exam, it is important that the type of questions with heavy memorization is avoided. Instead, the type of questions that check the students' ability to think creatively is needed. This type of assessment will be able to distinguish the ability of the students and therefore the grade can be distributed in a fair manner. The exam that contains standard and procedural type of questions are good to help the students who are weak. But if many people have excellent scores, it would be hard to distinguish between good students and those who are weak. On the other side of the spectrum, we should also avoid an exam with type of questions that are simply too challenging for everyone. This will in turn provide a very low average to all students and again we could not distinguish which students are good and which students are weak. An ideal exam would be when the result resembles the curve of normal distribution, with an average slightly around the middle and with not too large of standard deviation value.

In addition, a sufficient time should be provided to the students during the exam. The current exam time during the regular semesters (Fall and Spring) is only 60 minutes. Very often, the students need to answer around seven questions, where each question carries a different score but still comparable weight, totalling to a perfect score of 100. Within this short period of time, the students are forced to write their exam paper almost instantly, encouraging them to memorize the technique to solve problems as it were. However, if we would like to encourage our students to improve their creative thinking style, a sufficient time should be given for them to think, formulate and write down their answer during the exam. During exercise session in the class, we can prepare the students by training them with problem solving activities.

For Winter semester 2010, the midterm exam is also has a duration of 60 minutes with seven questions. Yet, the final exam consists of five questions, each question covers the corresponding chapter in the textbook and the exam duration is 120 minutes instead of one hour. Within this two hours, the students have ample time to think before they write down their answers. Providing a longer examination time also gives an opportunity for the instructor to present more challenging questions that require creative thinking skill on the side of the students. By doing so, I hope to separate the population based on the grade distribution. The type of the questions that tests the creative thinking should be non procedural, require an understanding and insight to combine several concepts that the students have learned in solving the problems. Designing this type of exam questions requires thorough effort. Furthermore, the questions should be `novel' in nature, for which the students have never seen before and yet within the level of the course and do not go beyond the scope of the material.

In fact, if we want to inculcate the ability of creative thinking in problem solving, we need to design the curriculum in such a way that it accommodates the need. See for instance Isaksen and Parnes (1985) who reported a survey of 150 curriculum planners about their knowledge, attitudes and behaviour regarding deliberate development of creative thinking and problem solving. A six-step creative problem solving model is also proposed in this article. Since creative thinking is closely related to problem solving, we also need to train our students to solve the problems effectively and present their answer in a communicative way. It is observed that many students just present merely the rough works instead of step by step explanation to certain mathematical problem, in this context is Calculus. This observation is captured when we asked the students to write on the whiteboard the solution of exercises selected from the book. Many students jump their steps since they might consider that the calculation is too simple, while other students just simply write symbols and formulas without having clear understanding of what those symbols mean.

## Exams

There are several options in designing the type of questions for the assessment of Calculus courses, depending on the need and the purpose of the assessment itself. The term `assessment' is generally used to refer to all activities that instructors use in helping the students to learn and to gauge their progress. So, in this context, midterm and final term exams are just two components of assessment. Depending on the type of questions, these exams can be categorized as either objective or subjective. Objective type is a form of question which has a single correct answer. On the other hand, subjective type is a form of question which may have more than one correct answer or as normally occurs in the context of Natural Sciences subjects, there are more than one way to express the correct answer. The objective type of questions includes true or false answers, multiple choice questions, multiple response and matching questions. The subjective type of questions includes extended response questions and essays. The former type of questions is very popular in the United States and is increasingly popular for computerized and online assessment formats. Many countries around the world have also adopted the objective type of questions at different educational levels.

A good quality examination should meet the criteria of high levels of reliability and validity. Nevertheless, these criteria may vary from one situation to the other. The reliability is related to the consistency of the exam itself. In other words, a reliable exam would consistently achieve similar results for the same or similar cohort of students. The validity is a measurement what is intended to measure. This means that the content of an exam should measure the stated objective in the syllabus or teaching plan. A correlation of score obtained to outside reference should be another factor. For instance, a good score in Calculus 1 should accurately predict a good score in Calculus 2 as well or in Engineering Mathematics or in Vector Calculus, or other advanced mathematics courses that need the basic understanding and strong foundation in Calculus. The exam should also correspond to other significant variables. For example, the students who major in Natural Science should not perform differently from the students who major in Engineering, or even in Economics, if they need to take Calculus courses.

With this in mind, the questions for every examination are carefully examined and checked. The remark that we would to mention over here is that it is very essential for an exam paper to be moderated by someone else who is knowledgeable in the subject or in the field before it goes into the printing process. In this context, the course coordinator has the responsibility to assure the reliability and the validity of each question. Thus, making a good quality of exam paper. The coordinator receives questions from several instructors. Combining these questions need a thorough screening, so that the exam paper gives a good representation of the syllabus covered and also a coherent structure from one question to the other. It is not only the questions that need to be checked, but also the solutions, to make sure that it is free from any mistakes or typographical errors. For Winter semester 2010, since there is no such coordinator, the checking is done several times by the instructor and the final version is given to the TA to be checked as well. A sufficient effort should be taken to minimize the possibility of mistakes in the exam questions of the answer parts. Another small remark. During the exam, it may be necessary to re-arrange the position of the students so that they do not sit too close with each other. A sufficient distance should be provided to avoid distraction from their neighbours.

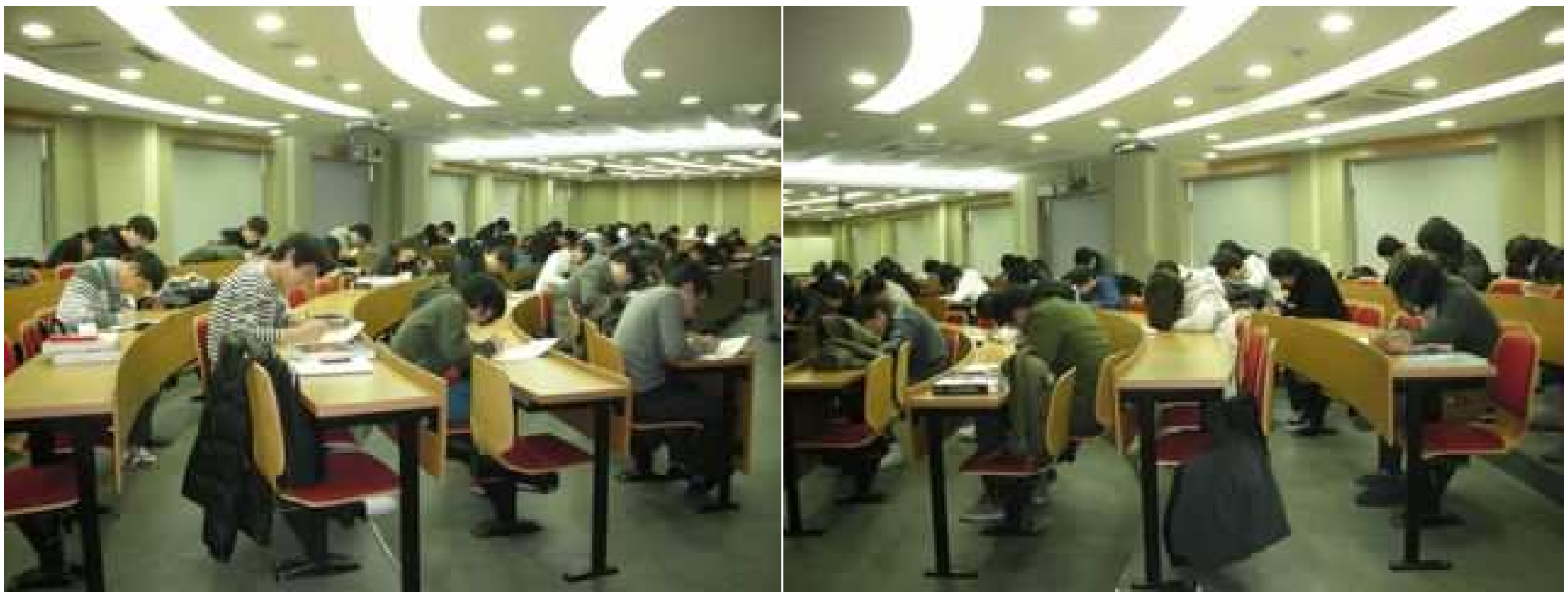

*Seating arrangement during the examination. The students are not sitting too close to their peers, but a reasonable distance should be provided.*

# Grading

Since the students need to follow the class in English, it is expected that the grading is more generous than the course in local, native Korean. One student remarked at The Sungkyun Times November 2010 edition that he is admitted to the university not because of his English ability, but because of his capabilities in Natural Sciences and Mathematics subjects. Therefore, there are a strong expectation from the students' side that if they follow the course in English instead of in Korean, they will obtain an easier grade for the sake of understanding the language. However, the Mathematics Department at SKKU has a different view regarding the grade distribution for Basic Science and Mathematics modules. The following tables provides some guidelines of grading for both undergraduate and graduate courses offered by the Math Department. These guidelines are particularly practical for Summer and Winter semesters, where there is no course coordinator is assigned to supervise the classes. However, these are only rough guidelines, and a flexibility is allowed within 3% difference from one class to the other. Big differences should be avoided since it may present a basis for claims by the students. The guidelines are effective per June 2010. By following these guidelines, we hope that the quality of our students will be maintained at the high level.

| Grade | The number of students (percentage) | Accumulative percentage |
|---|---|---|
| A+ | 12% | 12% |
| A | 13% | 25% |
| B+ | 15% | 40% |
| B | 15% | 55% |
| C+ | 10% | 65% |
| C | 10% | 75% |
| D+/D | 15% | 90% |
| F | 10% | 100% |

*Grading guideline for all BSM classes, practically implemented during the Summer and Winter semesters.*

| Grade | The number of students (percentage) | Accumulative percentage |
|---|---|---|
| A+ | 15% | 15% |
| A | 15% | 30% |
| B+ | 15% | 45% |
| B | 20% | 65% |
| C+ | 10% | 75% |
| C | 10% | 85% |
| D+/D | 5% | 90% |
| F | 10% | 100% |

*Grading guideline for all undergraduate classes. Particularly implemented during the regular semesters.*

| Grade | The number of students (percentage) | Accumulative percentage |
|---|---|---|
| A+ | 30% | 30% |
| A | 30% | 60% |
| B+/B | 25% | 85% |
| C+/C/D+/D/F | 15% | 100% |

*Grading guideline for all graduate classes.*

# Teaching Evaluation

## Questionnaires

In order to obtain students' feedback on the implemented teaching approach, a questionnaire is distributed to the students during the final exam of Winter 2010 semester. The questionnaire contains 20 questions and a space is provided for students to give their personal additional comments if they wish. It may take around three to five minutes to fill it out. The questions range from general to additional type of questions, including teaching strategies, students learning, assessment, professional attitude and teaching resources. The students may give a rate to each of these questions, ranging from 1 (the best) and 5 (the worst). A 3 is considered a neutral position. The sample of this questionnaire can be found below. Out of 78 distributed questionnaire papers, only 57 returned questionnaires are obtained with two invalid data. The following is the corresponding obtained score and the percentage of those who opted scores 1 and 2.

From this result we could observe that generally, more than half of the students give positive responses to all items in the questionnaire. In particular, the items `helpful attitude toward students’ and `treating students fairly and respectfully’ show good scores of less than 2.0 with the percentage of the students who opted for strongly agree and agree to these statements are 67% and 69%, respectively. On the other hand, the items `sensitivity to cultural difference’ and `encouraging the students to work as part of a team’ receive a rather bad score of 2.47 and 2.39, respectively. The percentage of the students who strongly agree and agree to these items are around 50%. Thus, there is definitely still a large room for improvement in this areas. When processing the questionnaire, we noticed that many students rather confuse with the statements on sensitivity to cultural difference. Perhaps the statement is not carefully written in proper English. Therefore, it is possible for next time that we should reword the sentence carefully or simply remove the item from the list. As far as we concerned, we have tried our best to encourage the students to work as part of a team. As we mentioned earlier, a suggested group work has been formed based on the result of the diagnostic test. Furthermore, a freedom is given to the students to choose and form their own group according to their wish when completing assignments. So, with this remarks, it is good for us as instructors to see things work out from the students’ perspective and instead of ours. Yes, the effort to encourage the students to work as part of a team only halfway successful. Definitely, we still could try some other ways to improve this area in the future.

| **Topical questions** | **Score** | **Percentage** |
|---|---|---|
| Course organization | 2.05 | 65% |
| Excellence in communicating ideas and information | 2.07 | 63% |
| Stimulating students' interest | 2.23 | 54% |
| Helpful attitude toward students | 1.93 | 67% |
| Effectiveness in teaching | 2.02 | 65% |
| Effectiveness of each session | 2.18 | 61% |
| Regularity in using examples and illustrations | 2.02 | 64% |
| Encouraging students' participation | 2.14 | 64% |
| Encouraging to work as part of a team | 2.39 | 50% |
| Creating comfortable learning environment | 2.11 | 61% |
| Balance between teacher and students | 2.11 | 55% |
| Effectiveness in developing critical and analytical skills | 2.21 | 55% |
| Success in improving to work independently | 2.21 | 55% |
| Success in helping students to learn how to learn | 2.18 | 57% |
| Improving students' communication skills | 2.34 | 52% |
| Reasonableness in workload | 2.09 | 62% |
| Accessibility to students | 2.11 | 60% |
| Sensitivity to cultural difference | 2.47 | 51% |
| Treating students fairly and with respectfully | 1.95 | 69% |
| Using appropriate resources | 2.09 | 66% |

*Result from the distributed questionnaire. The score indicates the obtained average, and ranges from 1 (best) and 5 (worst). The percentages indicate the number of students who opted scores 1 and 2.*

**Student Teaching Evaluation Form**

**GEDB002-41 Calculus 2 Winter 2010**

**Generic Questions**

| | | | | |
|---|---|---|---|---|
| 1. | How organised have you found Dr. Karjanto's contribution to this course? | Well organised | 1 2 3 4 5 | Disorganised |
| 2. | How would you rate Dr. Karjanto's ability to communicate ideas and information? | Excellent | 1 2 3 4 5 | Poor |
| 3. | How much has Dr. Karjanto stimulated your interest in the subject? | Very much | 1 2 3 4 5 | Not at all |
| 4. | How would you describe Dr. Karjanto's attitude toward students in this course? | Very helpful | 1 2 3 4 5 | Not at all helpful |
| 5. | Overall, how effective have you found Dr. Karjanto in teaching this course? | Very effective | 1 2 3 4 5 | Not at all effective |

**Additional Questions**

**Teaching Strategies**

| | | | | |
|---|---|---|---|---|
| 6. | How effectively did Dr. Karjanto structure each session? | Very effectively | 1 2 3 4 5 | Not at all effectively |
| 7. | Did Dr. Karjanto make good use of examples, illustrations, or other techniques to explain difficult concepts? | Regularly | 1 2 3 4 5 | Rarely |
| 8. | How successful was Dr. Karjanto in encouraging your participation? | Very successful | 1 2 3 4 5 | Not at all successful |
| 9. | How successful was Dr. Karjanto in encouraging you to work as part of a team? | Very successful | 1 2 3 4 5 | Not at all successful |
| 10. | Did Dr. Karjanto create a learning environment in which you felt comfortable? | Very much so | 1 2 3 4 5 | Not at all |
| 11. | Did Dr. Karjanto achieve a good balance between teacher contribution and student participation? | Very good | 1 2 3 4 5 | Poor |

**Student Learning**

| | | | | |
|---|---|---|---|---|
| 12. | How effective was Dr. Karjanto in helping you to develop your critical and analytical skills? | Very effective | 1 2 3 4 5 | Not at all effective |
| 13. | How successful was Dr. Karjanto in helping you to improve your ability to work independently? | Very successful | 1 2 3 4 5 | Not at all successful |
| 14. | How successful was Dr. Karjanto in helping you to learn how to learn? | Very successful | 1 2 3 4 5 | Not at all successful |
| 15. | Did Dr. Karjanto help you to improve your communication skills? | Definitely | 1 2 3 4 5 | Not at all |

**Student Teaching Evaluation Form**

**GEDB002-41 Calculus 2 Winter 2010**

**Assessment**

| | | | | |
|---|---|---|---|---|
| 16. | Did you find the workload required in Dr. Karjanto's course / section of the course reasonable? | Very reasonable | 1 2 3 4 5 | Not at all reasonable |

**Professional Attitudes**

| | | | | |
|---|---|---|---|---|
| 17. | How accessible was Dr. Karjanto to students? | Very accessible | 1 2 3 4 5 | Not at all accessible |
| 18. | How sensitive was Dr. Karjanto to cultural differences? | Very sensitive | 1 2 3 4 5 | Not at all sensitive |
| 19. | Did Dr. Karjanto treat students fairly and with respect? | Always | 1 2 3 4 5 | Seldom |

**Resources**

| | | | | |
|---|---|---|---|---|
| 20. | Did Dr. Karjanto use appropriate resources (print, OHTs, videos, CD-ROM, etc.) to enhance your understanding of this course? | Definitely | 1 2 3 4 5 | Not at all |

**Your Additional Comments:**

*The questionnaire distributed to the students during Winter semester 2010. It consists of 20 questions with five options given and a space is provided for students to give their additional comments.*

## Students' feedback in ASIS

The students' feedback obtained from ASIS so far is only available for the Fall 2010 semester, but not available for the Winter 2010 semester yet. From the feedback received, we observed that generally the students give a positive feedback on the teaching. Many students simply expressed their gratitude for the class. Some of them also give positive commendations. Yet, the comments from some other students are simply `no comment'. The percentile scores for two Calculus courses taught during Fall 2010 semester are 93% and 91%. So, overall, there is a positive response from the students. The questionnaire consists of 10 items, given as follows:

- The course was conducted according to the course syllabus provided to students at class registration.
- The instruction was carried out in the language stated in the syllabus.
- The instructor's explanations were clear and understandable.
- The instructor provided sufficient guidelines for students in preparing presentations and/or written assignments in the designated language.
- The instructor was available to students asking questions.
- The course materials (e.g. readings, lecture notes, in-class exercises) contributed to

learning the subject matter.
- This course improved my understanding of concepts and principles in this field.
- The required readings, assignments, and other preparatory materials were sufficient.
- The overall quality of this course was satisfactory.
- There were no class cancellations due to the instructor's personal matters. If classes had to be cancelled due to unavoidable reasons, the instructor held sufficient make-up classes.

The result of the students’ evaluation is shown below.

## 2010 학년도 제 2학기 개인별 강의평가결과 통보서

**소속 :** 학부대학 **성명 :** 나타나엘칼잔토 **직급 :** (외)조교수

귀하께서 2010 학년도 제 2학기에 담당하신 교과목에 대한 강의평가결과를 아래와 같이 알려드리니 다음 학기 강의준비에 참고하시기 바랍니다.

■ 담당과목별 강의평가 결과

| 교 과 목 명 | 강의평가유형 | 학수번호 | 분반 | 차수 | 강좌유형 | 수강인원 | 참여인원 | 백분위점수 |
|---|---|---|---|---|---|---|---|---|
| 미분적분학2 | 국제어형 | GEDB002 | 49 | 2 | 이론과목 | 69 | 69 | 93 |
| 미분적분학2 | 국제어형 | GEDB002 | 56 | 2 | 이론과목 | 63 | 58 | 91 |
| 이산수학 | ABEEK형 | GEDB007 | 43 | 2 | 이론과목 | 75 | 72 | 89 |

**담당과목 전체 백분위 점수 : 91 점**

■문항별 응답인원

| 학수번호 | 분반 | 응답구분 | 문항1 | 문항2 | 문항3 | 문항4 | 문항5 | 문항6 | 문항7 | 문항8 | 문항9 | 문항10 | 계 |
|---|---|---|---|---|---|---|---|---|---|---|---|---|---|
| GEDB002 | 49 | 매우긍정 | 40 | 39 | 38 | 41 | 40 | 38 | 40 | 37 | 40 | 41 | 394 |
| | | 긍정 | 16 | 16 | 13 | 14 | 17 | 15 | 16 | 16 | 15 | 12 | 150 |
| | | 보통 | 12 | 13 | 15 | 13 | 11 | 15 | 12 | 15 | 13 | 15 | 134 |
| | | 부정 | 1 | 1 | 3 | 1 | 1 | 1 | 1 | 1 | 1 | 1 | 12 |
| | | 매우부정 | 0 | 0 | 0 | 0 | 0 | 0 | 0 | 0 | 0 | 0 | 0 |
| | | 무응답 | 0 | 0 | 0 | 0 | 0 | 0 | 0 | 0 | 0 | 0 | 0 |
| GEDB002 | 56 | 매우긍정 | 30 | 30 | 26 | 27 | 24 | 24 | 24 | 26 | 23 | 29 | 263 |
| | | 긍정 | 22 | 20 | 22 | 19 | 25 | 25 | 22 | 19 | 24 | 21 | 219 |
| | | 보통 | 6 | 8 | 9 | 11 | 9 | 7 | 12 | 12 | 10 | 8 | 92 |
| | | 부정 | 0 | 0 | 1 | 1 | 0 | 2 | 0 | 1 | 1 | 0 | 6 |
| | | 매우부정 | 0 | 0 | 0 | 0 | 0 | 0 | 0 | 0 | 0 | 0 | 0 |
| | | 무응답 | 5 | 5 | 5 | 5 | 5 | 5 | 5 | 5 | 5 | 5 | 50 |
| GEDB007 | 43 | 매우긍정 | 33 | 37 | 31 | 30 | 33 | 30 | 31 | 31 | 32 | 34 | 322 |
| | | 긍정 | 27 | 23 | 22 | 24 | 23 | 25 | 24 | 24 | 21 | 25 | 238 |
| | | 보통 | 10 | 10 | 15 | 13 | 12 | 13 | 15 | 14 | 16 | 10 | 128 |
| | | 부정 | 0 | 0 | 1 | 3 | 2 | 2 | 0 | 1 | 1 | 1 | 11 |
| | | 매우부정 | 2 | 2 | 3 | 2 | 2 | 2 | 2 | 2 | 2 | 2 | 21 |
| | | 무응답 | 2 | 2 | 2 | 2 | 2 | 2 | 2 | 2 | 2 | 2 | 20 |

*The result of students’ evaluation for two Calculus courses and one BSM course during Autumn 2010 semester.*

The followings are some comments from the students:
- Thank you.
- Nice.
- None/nothing.
- No recommendations.
- Everything was all right.

- No comment.
- Good.
- It was a perfect lecture.
- It’s great. I really like your teaching.
- It is a nice class.
- I understand your pronunciation.
- It was good and helpful.
- The homework was too little and the quiz was too easy.

## Peer observation reports

Peer observation in teaching activities is very essential for every instructor. It is a method of gaining feedback to improve our teaching skills. The rationale behind peer observation is that it offers insight regarding the improvement of teaching. Peer observation is the observation of instructors by instructors, usually, though not always, on a reciprocal basis. This activity is one of many possibilities for the college to share good practice in learning, teaching and assessment methods. By doing so, the members of the college are encouraged to discuss about their teaching with colleagues and to learn from each other’s practices. Peer observation is not about monitoring or judging teaching staff. Rather, it should be seen as a support system. The pairings in peer observation can be done between mentor and novice or experienced lecturer and experienced lecturer. For the former case, the focus is to help the novice to develop their teaching skills both by observing and being observed by an experience colleague. For the latter case, the objective is to provide opportunities for experienced lecturers to reflect on their teaching.

Whether we are new or experienced and whether we are observing or being observed, there are a number of advantages to be gained when we involve in peer observation. The following advantages are taken and adapted from the web page on peer observation of the University of Essex.

- By watching another colleague approaches a particular teaching situation can often help us to better understand how to approach similar situations in our own course.
- It can often prompt us to reconsider our own particular teaching approaches and perhaps adopt a different approach or refine our existing practice slightly.
- It provides an opportunity to talk about the problems we might encounter with a particular approach and a chance to explore possible solutions or alternative approaches with our colleagues who may have experienced similar difficulties.
- To receive feedback from more established colleagues and to share good practice from the incoming members of staff.
- When new approaches to improve our teaching sessions are implemented, both formal and discussions in a number of forums give us an opportunity to how something is actually working in practice.

- We can see a completely different direction on a particular teaching approach when we observe a class in different department.

For some examples of peer observation practices: please visit the websites of peer observation at the University of Essex, The University of Nottingham and Leeds Metropolitan University.

During Autumn 2010 semester, we have conducted peer observation activities. I observed Dr. Lahaye's class and acted as an observer and on another occasion, Dr. Lahaye came to my class and he acted as an observer. Both of us receive valuable feedback from each other as we acted as both instructor and observer. The report of our peer observation can be found in the Appendix of this report. The form is adapted from the University of Nottingham's peer observation form. With this opportunity, I would like suggest the Center of Learning and Teaching at SKKU to include the peer observation as a regular feature of the teaching activities at our university.

# Conclusions and Future Perspectives

From our study we have learned that students' learning improves when they are actively engaged with the study material, instead of only sitting passively in the classroom. More success can be achieved when the classroom activities are also fun. With a proper balance between lecturing students new concepts and activities in which students, alone or in groups, have to struggle themselves with these concepts, makes the learning time in the classroom more effective and the time spent in class becomes more enjoyable for the students. Students also show appreciation for this style of teaching. Despite classroom participation and group activities are generally more successful in classes with a small number of students, we have accomplished promising and good results even though the size of our classes was over 70 students.

Improving students' participation in Calculus classes has achieved a measure of success to a certain extent. Some students are willing to participate in class activities, but others are too shy. Because the students' educational backgrounds are diverse, especially the weaker students should discern how to benefit from classroom participation. As for the Korean students in our classes (the vast majority), the proficiency differs according to their high school education. For example, in General Physics the 70 students in one class are a mixture of four distinct levels: (1) the top rank students from the science high schools, (2) second rank students with high school Physics 2 level, (3) third rank students with high school Physics 1 level, and (4) students without any high school physics. The gap between (2) and (3) is surprisingly large, let alone the gap between (1) and (4). It is almost impossible to design a teaching style that suits all students. However, for class activities with groups or assignments in teams it might be worthwhile to take the proficiency level into account. When the proficiency gap within the group or team members is too large, then the best student(s) will most probably do most of the work in exchange for a 'compensational treat' by the other members, instead of teaching the other members and share the workload. Hence, it may work better if teams are formed with students of the same or similar level. Then the groups with top rank students will get the most challenging problems to solve, groups with the second rank students somewhat easier problems, etc. Before the formation of the teams, the instructor can determine the proficiency of each student from the students' high school education level or from the results of a diagnostic test.

Organizing the midterm and final exams in exactly the same manner and let each one contribute equally to the final grade, has serious flaws. For students who flounder on the midterm exam, it is a shear impossible task to undo this in the final exam and some of the students may just give up after the midterm. Alternatives could be to give a lower weight percentage to a shorter, one-hour midterm exam and a higher weight percentage to a longer, two-hour final exam. In addition, in a somewhat longer exam students can be challenged into deeper thinking and possibly to show more personal creativity. Improvements of the instructor's teaching methods will also affect the learning objectives for the students and hence will require a different style of examination.

Although embracing students' participation strategies in the teaching style is a indispensable pedagogy tool, the students psychology can easily throw a spanner in the works. If the instructor does not specify well-defined regulations for the students' rewards for participating in the activities, students would lose interest and their efforts would diminish as well. The regulations must clearly state what grade incentives students can earn for their participation. Students therefore need to know how this is evaluated and how it affects their grade. Especially students who perform below average seem to need these kind of incentives.

Peer observations are particularly useful as an internal tool for the improvement of academic education. Although peer observations are not a common practice at SKKU, it is very useful not only for new faculty members, but also to those who have been teaching for a long period of time already. Peer observations are a standard support system for teaching staff at numerous institutions of higher learning all over the world. Therefore we recommend to implement this also at SKKU, because it is essential to the improvement of the quality of teaching.

We will continue to improve practical methods that have worked out already to help the students learning more actively in both inside the classroom and out-of-class, such as solving problems in a group, presenting answers in front of the class, and giving our-of-class assignments to teams of students. We are convinced that students can benefit from these activities and therefore the frequency of afore mentioned activities should be increased in the future. Another option for continuous improvement is the teaching style as a whole, for example by introducing problem-based and self-guided learning as part of the course.

# Appendices

# Peer Teaching Evaluation Form

***A, B, and C must be completed by the <u>instructor</u> before the teaching session.***

***D and E must be completed by the <u>observer</u> during or after the teaching session.***

**A: Factual Information**

Name of Observer: Natanael Karjanto

Name of Instructor: Rob J.W.E. Lahaye

Department: University College

Course name & code: General Physics 2 - GEDB008-41

Semester – Year: Fall 2010

Type of session to be observed (e.g. lecture, supervision, tutorial, etc) Lecture

Any comments on the nature of student group (e.g. is the session mandatory/or elective participation)?

The course is mandatory for almost all students in the group.

About 85 % of the 70 students enrolled in this course are freshmen.

On average attendance is roughly 50 students.

Are there any special features about the class composition or any special learning needs?

The class is given in English language.

All students are non-native speakers, mostly Koreans with a few Malay and Chinese.

**B: Class Information**

Topic The basic concepts of Einstein's special theory of relativity

Session length 1 hours 15 minutes

Observation length 1 hours 15 minutes

Approximate number of students 45

Date of session December 6, 2010

**C: Learning Objectives**

What are the specific learning objectives of the session? (e.g. knowledge, key skills, cognitive skills, practical/professional skills and transferable skills).

The theory of relativity has counter-intuitive consequences.
It is important that students get a feeling for the basic concepts,
so that the relativistic ideas become less counter-intuitive.

## D: Strengths and Areas for improvement

| 'Prompt' Questions | Comments<br>Strengths | <br>Areas for improvement |
|---|---|---|
| 1. Are the learning objectives clear to all? | Yes, after following the lecture, the students would understand basic concepts of the theory of relativity. | – |
| 2. Is the session well planned and clearly organized? | Yes, all planned materials are covered within the time slot period. These includes explanation and also giving the students opportunity to solve problems. | – |
| 3. Are the teaching and learning methods used appropriate? | Yes, the lecturer implemented both lecturing method and problem-based learning method in a balanced way. | – |
| 4. Is the style of delivery and the pace appropriate? | Yes, the style of lecturing and group work for students is implemented in a balanced way. | – |
| 5. Are the students encouraged to learn actively and participate? | Yes, the lecturer posed some problems and the students are organized in groups to allow them to solve the problems and thus enhanced their learning skills. | – |
| 6. Is appropriate use made of the accommodation and learning resources? | Yes, the lecturer made use of both the whiteboard, screen projector, pictures and other relevant animations to help students understand. | – |
| 7. Is the content appropriate to the needs of the students (e.g. use of examples)? | Yes, examples are taken from everyday life. Some examples also require deep thinking from the students. | – |

## E: Summary

Please summarize the overall quality of the session in relation to the stated learning objectives.

Overall, the lecture is a very excellent one.
The lecturer is very keen to help the students to understand certain difficult physical concepts by explaining in a clear way and giving opportunity to the students to solve some relevant problems.
In this way, the lecturer could meet the learning objectives in a good way.

**Instructor’s comments:**

The instructor appreciates the stimulating feedback and thanks the observer for positive comments.

**Observer’s comments:**

Being the observer, I enjoyed this interesting class session since I could learn something. Apart from this, I also could learn how a good lecturer that meet the learning objective should be conducted in an appropriate manner.

It is no doubt that the university should give more teaching incentive to maintain  a good lecturer like this one.

**Instructor’s signature** Rob Lahaye **Date** January 18, 2011

**Observer’s signature** Natanael Karjanto **Date** January 18, 2011

# Peer Teaching Evaluation Form

***A, B, and C must be completed by the <u>instructor</u> before the teaching session.***

***D and E must be completed by the <u>observer</u> during or after the teaching session.***

**A: Factual Information**

Name of Observer: Rob J. W. E. Lahaye

Name of Instructor: Natanael Karjanto

Department: University College

Course name & code: Calculus - GEDB002-49

Semester – Year: Fall 2010

Type of session to be observed (e.g. lecture, supervision, tutorial, etc) Lecture

Any comments on the nature of student group (e.g. is the session mandatory/or elective participation)?

The session is mandatory for students majoring in Engineering and Natural Science.

Are there any special features about the class composition or any special learning needs?

The class is given in English for non native English speakers;
the majority of students are Korean with a few foreign students.

**B: Class Information**

Topic____________________

Session length 1 hours 15 minutes

Observation length 1 hours 15 minutes

Approximate number of students 20

Date of session December 9, 2010

**C: Learning Objectives**

What are the specific learning objectives of the session? (e.g. knowledge, key skills, cognitive skills, practical/professional skills and transferable skills).

The students are able to use Green's theorem when solving problems involving line integrals along a closed curve or double integrals over a domain enclosed by a closed curve.

## D: Strengths and Areas for improvement

| 'Prompt' Questions | Comments<br>Strengths | <br>Areas for improvements |
|---|---|---|
| 1. Are the learning objectives clear to all? | Objectives are clearly stated by using both whiteboard and computer screen. | Try to introduce new concepts by asking questions first and let students think for a while. |
| 2. Is the session well planned and clearly organized? | There is a good balance between lecturing new concepts and doing exercises with the new concepts. | Allow for a classroom discussion after a student solves a problem on the whiteboard. |
| 3. Are the teaching and learning methods used appropriate? | Yes. | When asking a question to the class, wait longer for students to organize their thinking. |
| 4. Is the style of delivery and the pace appropriate? | The pace and speech is slow enough for a non-native speaker audience. | Students remain very quiet in class; maybe using interesting challenges as questions may trigger more response? |
| 5. Are the students encouraged to learn actively and participate? | The instructor demands frequently response from the students and forces students to solve problems on the whiteboard. | Students are encouraged, but remain passive and are reluctant to participate. To change this needs structural changes of teaching style. |
| 6. Is appropriate use made of the accommodation and learning resources? | Whiteboard and computer projection of the textbook are used simultaneously. | Use the monitor screen pen to point at and add remarks into the computer projection. |
| 7. Is the content appropriate to the needs of the students (e.g. use of examples)? | Several examples clarify the newly learned theory. | Try to stimulate a classroom discussion about the obtained results after finishing each example. |

## E: Summary

Please summarize the overall quality of the session in relation to the stated learning objectives.

The chosen method for introducing new concepts is good. In this particular class the explanation of the Green's theory is well done by referring to the textbook via the computer projection screen and explaining it further on the whiteboard.

Improvements should be considered for the problem solving sessions.
When a student solves a problem on the whiteboard, let the class first go over the solution before the lecturer adds comments. Being told by another student that there is a mistake, has a milder impact than when the instructor makes corrections (do the latter only as a last resort, when nobody finds the mistakes).

**Instructor's comments:**

The comments from the observer is very useful and beneficial.
Some areas for improvement will be paid attention to and some positive actions are to be implemented.

The instructor thanks the observer for the valuable suggestions.

**Observer's comments:**

It is an interesting experience to observe this lecture.
Despite the initial passive attitude of the students,
the instructor is persistent to change this in the problem session.
Sitting in the back of the classroom myself, I could see that this enhances the attention and concentration of other students as well.
This is a very good and valuable quality of the instructor.

**Instructor's signature** Natanael Karjanto **Date** January 18, 2011

**Observer's signature** Rob Lahaye **Date** January 18, 2011